\documentclass[twocolumn,preprintnumbers,amssymb,amsmath,aps,floatfix,prc,nofootinbib,superscriptaddress]{revtex4}

\usepackage{epsfig}
\usepackage{bm}
\usepackage{amssymb}
\usepackage{amsmath}
\usepackage{color}
\usepackage{subfigure}
\usepackage[unicode=true,pdfusetitle,
 bookmarks=true,bookmarksnumbered=false,bookmarksopen=false,
 breaklinks=false,pdfborder={0 0 1},backref=false,colorlinks=true]
 {hyperref}
\hypersetup{
 pdfborderstyle=,linkcolor=blue,citecolor=cyan}

 \usepackage{ulem}

\begin{document}

\title{Local and global polarization of $\Lambda$ hyperons across RHIC-BES energies:\\ the roles of spin hall effect, initial condition and baryon diffusion}

\author{Xiang-Yu Wu}
\email{xiangyuwu@mails.ccnu.edu.cn}
\affiliation{Institute of Particle Physics and Key Laboratory of Quark and Lepton Physics (MOE), Central China Normal University, Wuhan, Hubei, 430079, China}

\author{Cong Yi}
\email{congyi@mail.ustc.edu.cn}
\affiliation{Department of Modern Physics, University of Science and Technology
of China, Hefei, Anhui 230026, China}

\author{Guang-You Qin}
\email{guangyou.qin@ccnu.edu.cn}
\affiliation{Institute of Particle Physics and Key Laboratory of Quark and Lepton Physics (MOE), Central China Normal University, Wuhan, Hubei, 430079, China}

\author{Shi Pu}
\email{shipu@ustc.edu.cn}

\affiliation{Department of Modern Physics, University of Science and Technology
of China, Hefei, Anhui 230026, China}

\begin{abstract}
We perform a systematic study on the local and global spin polarization
of $\Lambda$ and $\overline{\Lambda}$ hyperons in relativistic heavy-ion collisions at beam energy scan energies via the (3+1)-dimensional CLVisc hydrodynamics model with AMPT and SMASH initial conditions.
Following the quantum kinetic theory, we decompose the polarization vector as the parts induced by thermal vorticity, shear tensor and the spin Hall effect (SHE). We find that the polarization induced by SHE and the total polarization strongly depends on the initial conditions.
At $7.7$GeV, SHE gives a sizeable contribution and even flips the sign of the local polarization along the beam direction for AMPT initial condition, which is not observed for SMASH initial condition. Meanwhile, the local polarization along the out-of-plane direction induced by SHE with AMPT initial condition does not always
increase with decreasing collision energies. Next, we find that the
polarization along the beam direction is sensitive to the baryon diffusion coefficient, but the local polarization along the out-of-plane direction is not.
Our results for the global polarization of $\Lambda$ and $\overline{\Lambda}$ agree well with the STAR data.
Interestingly, the global polarization of $\overline{\Lambda}$ is not always larger than that of $\Lambda$ due to various competing effects.
Our findings are helpful for understanding the polarization phenomenon and the detailed structure of quark-gluon plasma in relativistic heavy-ion collisions.

\end{abstract}
\maketitle

\section{Introduction}

The most vortical fluid in nature, whose averaged angular velocity is as large as $\omega \sim 10^{22}~s^{-1}$, has been discovered by STAR Collaboration at the Relativistic Heavy-Ion Collider (RHIC) \citep{STAR:2017ckg,STAR:2021beb}.
Such a large vorticity originates from the huge initial orbital angular momentum carried by Quark Gluon Plasma (QGP) in non-central heavy-ion collisions.
Due to the spin-orbit coupling, the rapidly rotational fluid can induce the %global
spin polarization of the emitted hadrons, perpendicular to the reaction plane \citep{Liang:2004ph,Liang:2004xn,Gao:2007bc}.
In addition, non-central heavy-ion collisions can create an anisotropic QGP fireball with non-Bjorken expansion at the initial stage in heavy-ion collisions.
The inhomogeneous expansion of the anisotropic QGP fireball can lead to a periodic azimuthal angle dependence of the local spin polarization along both transverse and longitudinal directions  \citep{Becattini:2017gcx,Xia:2018tes, STAR:2019erd,Lisa:2021zkj}.
Compared to anisotropic collective flows, the global or local spin polarizations involve more details of the QGP, such as the gradients of chemical potential, temperature and flow velocity,
therefore provide a novel probe to the detailed structure and transport properties of QGP produced in high-energy nuclear collisions.

The spin polarization phenomena in the relativistic heavy ion collisions have been extensively studied in many aspects. The vorticities generated after the collisions are studied by various hydrodynamic models \cite{Betz:2007kg,Csernai:2013bqa,Becattini:2013vja,Becattini:2015ska,Pang:2016igs,Alzhrani:2022dpi} and
transport models \cite{Jiang:2016woz,Deng:2016gyh,Li:2017slc,Wei:2018zfb}.
Also viscous hydrodynamics models \citep{Ryu:2021lnx,Yi:2021ryh,Yi:2021unq,Fu:2022myl,Fu:2021pok,Fu:2020oxj,Pang:2016igs,Wu:2019eyi,Becattini:2016gvu,Karpenko:2016jyx,Zhao:2017rgg, Qin:2013bha, Schenke:2014zha, Bzdak:2014dia, Zhao:2019ehg,Schenke:2021mxx,Bozek:2011if,Bozek:2013uha,Bozek:2016kpf,Nagle:2018nvi}
and transport models \citep{Wei:2018zfb,Li:2017slc,Xia:2018tes,Vitiuk:2019rfv} have been very successful to reproduce the increasing global polarization of $\Lambda$ and  $\bar{\Lambda}$  with decreasing collision energies $\sqrt{s_{NN}}$
for over a wide range of collision energies ($7.7-200$~GeV) via performing the modified Cooper-Frye formula \citep{Becattini:2013fla,Fang:2016vpj} (also see the other early pioneer work for the relativistic spinning
particles  by quantum statistical models \cite{Becattini:2007nd,Becattini:2007sr,Becattini:2013fla,Becattini:2016gvu}).

Recently, the experimental data at low collision energies \cite{STAR:2021beb,Kornas:2020qzi} can be partially described by hydrodynamical models \cite{Guo:2021udq,Ivanov:2020udj} and transport
models \cite{Deng:2020ygd,Deng:2021miw}.
Although several phenomenological models with external assumptions \citep{Liu:2019krs,Wu:2019eyi,Voloshin:2017kqp,Fu:2021pok,Becattini:2021iol} can describe the experimental data qualitatively, most hydrodynamics simulations, transport models and feed-down effects \citep{Becattini:2017gcx,Xia:2018tes, Becattini:2019ntv,Xia:2019fjf, Fu:2020oxj, Li:2021jvn} failed to describe the azimuthal angle dependence of the local spin polarization.
In particular, an opposite ``sign" is obtained from hydrodynamics and transport model calculations as compared to the experimental data.
More discussions and details can be found in recent reviews \citep{Wang:2017jpl,Florkowski:2018fap,Becattini:2020ngo,Becattini:2020sww,Gao:2020vbh,Liu:2020ymh, Hidaka:2022dmn}.

One reason to explain this mismatch is that the theoretical calculation and the numerical simulation only consider the thermal vorticity under the global equilibrium condition.
Recent studies \citep{Liu:2020dxg,Liu:2021uhn,Becattini:2021suc} have shown that beyond global equilibrium, the polarization induced by shear tensor (SIP) is helpful to solve the ``sign" problem for the local spin polarization (see also the early derivation of such effects for the massless fermions \citep{Hidaka:2017auj} and Refs.~\citep{Liu:2021nyg,Florkowski:2021xvy} for related studies). The numerical simulations with the help of the SIP for the $s$ quarks \cite{Fu:2021pok} or in the iso-thermal approximation \cite{Becattini:2021iol} result in quantitative agreement with experimental data by tuning appropriate parameters.
Soon, several studies from different groups find that the results are very sensitive to the equation of state and the temperature
gradient in hydrodynamical simulations \citep{Yi:2021ryh, Florkowski:2021xvy,Sun:2021nsg}. Meanwhile, these off-equilibrium effects have also been extended to the helicity polarization \citep{Yi:2021unq}, which may also be helpful for detecting the initial axial chemical potential \citep{Becattini:2020xbh,Gao:2021rom}.

In general, there are two possible ways to add the spin degrees of freedom to the system.
One macroscopic way is the
relativistic spin hydrodynamics \citep{Florkowski:2017dyn,Florkowski:2017ruc,Florkowski:2018ahw,Florkowski:2018myy,Florkowski:2019qdp,Florkowski:2019voj,Becattini:2018duy,Bhadury:2020puc, Hattori:2019lfp,Fukushima:2020qta,Fukushima:2020ucl,Li:2020eon,She:2021lhe,Montenegro:2017lvf,Montenegro:2017rbu,Florkowski:2018fap,Yang:2018lew,Shi:2020qrx,Gallegos:2021bzp,Hongo:2021ona}.
Although the analytic solutions for spin hydrodynamics in the simplified Bjorken \citep{Wang:2021ngp} and Gubser flows \citep{Wang:2021wqq} have been found, it is still challenging to numerically solve the spin hydrodynamics equations for relativistic heavy-ion collisions.
The microscopic description for the massive fermions is the quantum kinetic theory (QKT) \citep{Gao:2019znl,Weickgenannt:2019dks,Weickgenannt:2020aaf,Hattori:2019ahi,Yang:2020hri,Liu:2020flb,Weickgenannt:2021cuo,Sheng:2021kfc,Wang:2021qnt,Huang:2020wrr,Wang:2020dws,Wang:2019moi}, %with local equilibrium conditions \citep{Liu:2021uhn,Hidaka:2017auj}.
which is an extension of the chiral kinetic theory \citep{Son:2012wh,Son:2012zy,Stephanov:2012ki,Gao:2012ix,Chen:2012ca,Chen:2013iga,Chen:2014cla,Chen:2015gta,Hidaka:2016yjf}.
It manifests the extra correction terms to the spin polarization besides the thermal vorticity.
Recently, there are many studies on the interaction effects, e.g. the generic form of collison terms based on the Kadanoff-Baym equation  \citep{Hidaka:2016yjf,Yang:2020hri,Sheng:2021kfc}, the discussion on the non-local collisions \citep{Zhang:2019xya,Weickgenannt:2020aaf,Weickgenannt:2021cuo}, the simplified collision term based on the hard-thermal-loop approximation \citep{Fang:2022} and other related studies \citep{Sun:2017xhx,Li:2019qkf,Kapusta:2019sad,Carignano:2019zsh,Liu:2019krs,Bhadury:2020puc,Hou:2020mqp,Fauth:2021nwe,Lin:2021mvw, Carignano:2021zhu}.
More references and related works can be found in the recent reviews for QKT \citep{Gao:2020vbh, Gao:2020pfu, Hidaka:2022dmn}.

Very recently, the polarization induced by the gradient of baryon chemical potential over temperature, also called spin Hall effect (SHE) in some works, has drawn a lot of attention. This effect has been studied in the early works for the massless fermions \citep{Son:2012zy,Chen:2016xtg, Hidaka:2017auj} and been extended to the massive fermions \citep{Fu:2022myl, Yi:2021ryh}.
In recent papers \citep{Fu:2022myl},
it is found that the polarization induced by the SHE has positive and significant contribution to the local polarization in baryon-rich region at lower collisions. The global polarization in the presence of SHE has been studied in Ref. \citep{Ryu:2021lnx, Alzhrani:2022dpi}. So far, several aspects of polarization induced by SHE is still missing, e.g. the dependence of initial conditions and the effects of baryon diffusion.

In this paper, we perform a systematic study on the local and global spin polarization of $\Lambda$ and $\overline{\Lambda}$ hyperons in Au+Au collisions across BES energies.
To provide a comprehensive understanding of the polarization,
in this work we concentrate on the studies of the polarization induced by SHE and the dependence of initial conditions and baryon diffusion. We utilize (3+1)-dimensional CLVisc hydrodynamics framework \cite{Pang:2012he, Pang:2018zzo, Wu:2021fjf}
with the AMPT (A-Multi-Phase-Transport) initial model \citep{Lin:2004en, Wu:2021hkv, Wu:2018cpc, Zhao:2017yhj}
and SMASH (Simulating Many Accelerated Strongly-interacting Hadrons) initial condition \citep{Weil:2016zrk,Schafer:2019edr,Mohs:2019iee,Hammelmann:2019vwd,Mohs:2020awg,Schafer:2021csj,Inghirami:2022afu} to simulate the dynamical evolution of QGP fireball. The SMASH initial condition, which includes the thickness effects of the nucleus, may provide extra useful information of QGP fireball in low energy region.

We will first show the local polarization along both beam and out-of-plane directions induced by different sources as a function of azimuthal angle. Next, we present the results for the total local polarization of both $\Lambda$ and $\overline{\Lambda}$. We analyze the contributions from SHE and discuss the dependence on initial conditions. Then, we focus on the effects of baryon diffusion. Since the gradient of baryon chemical potential, or SHE, is directly related to the baryon diffusion, we expect to observe the strong dependence on baryon diffusion. At last, we also plot the results for the global polarization of both $\Lambda$ and $\overline{\Lambda}$ hyperons across the BES energies.

The paper is organized as follows.
In Sec.\ref{model}, we first present the setup of (3+1)-dimensional viscous hydrodynamics model CLVisc with AMPT and SMASH initial conditions.
Then we briefly introduce the formula for the spin vector for different sources.
In Sec. \ref{results}, we show our numerical results for local and global polarizations and study the dependence of initial conditions and baryon diffusion.
Sec. \ref{summary} contains our summary and outlook.

\section{(3+1)-dimensional hydrodynamics CLVisc Framework at finite net baryon density}
\label{model}

In this section, we introduce the theoretical framework and setup for the numerical simulations. In Sec. \ref{subsec:initial}, we introduce two different initial conditions. Next, we present the relativistic (3+1)D CLVisc hydrodynamics framework and the values of parameters in Sec. \ref{subsec:HydroFrame}. We then review the particlization method used in the current studies in Sec. \ref{subsec:particlization}. At last, we show the general expressions for the polarization vector in Sec. \ref{subsec:SpinPolarization}.

\subsection{Initial condition} \label{subsec:initial}

The AMPT model has been widely used as the initial condition for hydrodynamical evolution in heavy-ion collisions at the LHC energies and the top RHIC energies \citep{Lin:2004en,Wu:2021hkv,Wu:2018cpc,Zhao:2017yhj,Zhao:2021vmu,Pang:2015zrq,Pang:2016igs}.
In AMPT model, the initial patrons are first produced via hard, semi-hard scattering and excited strings in HIJING model \citep{Gyulassy:1994ew,Wang:1991hta}.
Then the space-time evolution of the patrons are described via elastic scattering within Zhang's Parton Cascade (ZPC) model \citep{Zhang:1997ej} until they reach the iso-$\tau_0 $ hypersurface.

Alternatively, SMASH model, as a novel and modern hadronic transport approach, is developed to describe the non-equilibrium microscopic motions of hadrons at low energy heavy-ion collisions \citep{Weil:2016zrk,Schafer:2019edr,Mohs:2019iee,Hammelmann:2019vwd,Mohs:2020awg,Schafer:2021csj,Inghirami:2022afu}. It solves the relativistic Boltzmann equation effectively
\begin{equation}
  p^\mu \partial_\mu f + m F^\mu\partial_{p_\mu} ( f)= C[f],
  \label{eq:Boltzmann}
\end{equation}
where
$f(t,\mathbf{x},\mathbf{p})$ denotes one-particle distribution function, $p_\mu$ indicates the four-momentum of a particle, $m$ is the mass of particles and $F^{\mu}$ is the effective force induced by external mean-field potentials.
The collision kernel $C[f]$ includes elastic collisions, resonance formation and decays, string fragmentation for all mesons and baryons up to mass $\sim$ 2.35 GeV.
As for the resonances, the Breit-Wigner spectral functions with mass-dependent widths are utilized.
SMASH model also takes into account high-energy hadronic interactions according to the string model in Pythia 8 \citep{Sjostrand:2006za,Sjostrand:2007gs}.

After initializing hadrons via  Woods-Saxon distribution and Fermi motion, SMASH model describes the transport evolution of hadrons via scattering and interactions until they approach the hypersurface at iso-$\tau_0$ proper time and finally
outputs hadrons for hydrodynamics evolution.
In current work, we implement the SMASH-2.0 model and neglect the effect of the mean-field potentials for simplicity.

Here, we briefly comment on two models. In the AMPT model, all initial partons are assumed to be produced at $z=0$ plane when $t=0$, which is a good approximation at high energy collisions. On the other hand, at baryon rich region in BES energies, the finite size effect of the initial nucleus in the longitudinal direction  may not be negligible. This finite thickness effect has been considered in the SMASH model.

At the initial proper time $\tau_0$,  we assume that the patrons from AMPT model or hadrons from SMASH model reach the local thermal equilibrium.
The initial energy-momentum tensor $T^{\mu\nu}$ and the initial baryon current $J^{\mu}$ can be constructed at Melin coordinate:
\begin{eqnarray}
T^{\mu\nu} (\tau_0,x,y,\eta_s)&= K \sum\limits_{i} \frac{p^{\mu}_ip^{\nu}_i}{p^{\tau}_i} G(\tau_0,x,y,\eta_s), \label{eq:EMT_01} \\
J^{\mu} (\tau_0,x,y,\eta_s)&= \sum\limits_i Q_i \frac{p^{\mu}_i}{p^{\tau}_i} G(\tau_0,x,y,\eta_s),
\end{eqnarray}
where $G(\tau_0, x,y,\eta_s)$ denotes the Gaussian smearing
\begin{eqnarray}
&&G(\tau_0,x,y,\eta_s) \notag \\& = & \frac{1}{\mathcal{N}} \exp\left[-\frac{(x-x_i)^2+(y-y_i)^2}{2\sigma_r^2}- \frac{(\eta_s-\eta_{si})^2}{2\sigma^2_{\eta_s}}\right]. \label{eq:G_smearing}
\end{eqnarray}
Here
\begin{equation}
   p^{\mu}=\left[m_{T} \cosh \left(Y-\eta_{s}\right), p_{x}, p_{y}, \frac{1}{\tau_{0}} m_{T} \sinh \left(Y-\eta_{s}\right)\right], \nonumber
\end{equation}
is the four-momentum of hadrons or patrons with transverse mass $m_T$,
rapidity $Y$ and space-time rapidity $\eta_s$.
$Q$ indicates the  baryon charge for particles.
$\mathcal{N}$ is the normalization factor to keep the net baryon number conservation.
The overall normalization parameter $K$ and the gaussian smearing width $\sigma_r,\sigma_{\eta_s}$ can be determined
via comparing the hydrodynamics calculation to the experimental data on charged hadrons yield at the most central collisions \citep{Adamczyk:2017nof}.

In current work, we have averaged $5000$ fluctuating $T^{\tau\mu}$ and $J^{\tau}$ from initial events based on the AMPT or SMASH models to generate smooth initial distributions in given centrality bins,
which are determined by the number of initial patrons or impact parameter for the AMPT and SMASH models, respectively.

\begin{table*}[t]
\setlength{\tabcolsep}{10pt}
\centering
\caption{The parameter values for CLVisc hydrodynamics simulation with AMPT and SMASH initial conditions.The normalization factor $K$ and parameters for the Gaussian smearing $\sigma_r, \sigma_{\eta_s}$ are introduced in Eqs.~(\ref{eq:EMT_01},\ref{eq:G_smearing}). The $\tau_0$ is the initial time. The $C_{\eta_v}$ is specific shear viscoity defined in Eq.(\ref{eq:C_shear}). }
\resizebox{\textwidth}{1.2cm}{
\begin{tabular}{c| c c c c c | c c c c c}
\hline
& \multicolumn{5}{c|}{AMPT model} & \multicolumn{5}{c}{SMASH model}  \\
\hline \hline
$\sqrt{s_{NN}}$ [GeV] &K & $\tau_0$ [fm] & $\sigma_r$ [fm] & $\sigma_{\eta_s}$ & $C_{\eta_v}$
& K & $\tau_0$ [fm] & $\sigma_r$ [fm] & $\sigma_{\eta_s}$ & $C_{\eta_v}$  \\ \hline
7.7  & 1.4 & 2.0 & 1.0 & 0.7  & 0.2 & 1.0 & 3.2 & 1.0 & 0.35  & 0.2  \\ \hline
27   & 1.8 & 1.0 & 1.0 & 0.5  & 0.12 & 1.0 & 1.0 & 1.0 & 0.35  & 0.12  \\ \hline
62.4 & 1.7 & 0.7 & 0.6 & 0.55 & 0.08 & 1.0 & 0.7 & 1.0 & 0.55 & 0.08  \\ \hline
\end{tabular}
}
\label{table:amptparameter}
\end{table*}

\subsection{(3+1)D CLVisc hydrodynamics framework} \label{subsec:HydroFrame}

In this work, we implement the (3+1) dimensional CLVisc hydrodynamics framework to simulate the dynamical evolution of QGP.
Due to the finite net baryon density at RHIC-BES energies, we consider both the energy-momentum and baryon number conservation,
\begin{align}
&\nabla_{\mu} T^{\mu\nu}=0 \, ,\\
&\nabla_{\mu} J^{\mu}=0  \, ,
\end{align}
where $\nabla_{\mu}$ represents the covariant derivative operator in the Melin coordinate.
The energy-momentum tensor $T^{\mu\nu}$ and net baryon current $J^{\mu}$ can be decomposed into the following form:
\begin{align}
&T^{\mu\nu} = eU^{\mu}U^{\nu} - P\Delta^{\mu\nu} + \pi^{\mu\nu}\,, \\	
&J^{\mu} = nU^{\mu}+V^{\mu}\,,
\end{align}
where $e,P,n, U^{\mu}, \pi^{\mu\nu}, V^{\mu}$ are the energy density, pressure, net baryon density, the flow velocity, the shear-stress tensor and baryon diffusion current, respectively. The $\Delta^{\mu\nu} = g^{\mu\nu} - U^{\mu}U^{\nu}$ is a projection.
For simplicity, we neglect the effect of bulk viscosity \citep{Ryu:2015vwa,Ryu:2017qzn}. %\citep{Ryu:2015vwa,Ryu:2017qzn,Yi:2021ryh}.

The evolution of the dissipative currents $\pi^{\mu\nu}$ and $V^{\mu}$ is described by the following equations based on the Israel-Stewart second order hydrodynamics \citep{Denicol:2018wdp}:
\begin{align}
\Delta^{\mu\nu}_{\alpha\beta} (u\cdot \partial) \pi^{\alpha\beta} = &
 -\frac{1}{\tau_{\pi}}\left(\pi^{\mu\nu} - \eta_v\sigma^{\mu\nu}\right)
- \frac{4}{3}\pi^{\mu\nu}\theta
\nonumber
\\
&
-\frac{5}{7}\pi^{\alpha<\mu}\sigma_{\alpha}^{\nu>}+ \frac{9}{70}\frac{4}{e+P}\pi^{<\mu}_{\alpha}\pi^{\nu>\alpha}\,,
\nonumber
\\
\Delta^{\mu\nu} (u\cdot \partial) V_{\nu}  = &  - \frac{1}{\tau_V}\left(V^{\mu}-\kappa_B\bigtriangledown^{\mu}\frac{\mu_B}{T}\right)-V^{\mu}\theta
\nonumber \\
&-\frac{3}{10}V_{\nu}\sigma^{\mu\nu}\,,
\end{align}
where $\theta = \partial \cdot u$ is the expansion rate, $\sigma^{\mu\nu} = \partial^{<\mu} u^{\nu>}$ is the symmetric shear tensor,
$\eta_v$ and $\kappa_B$ are shear viscosity and baryon diffusion coefficient, respectively.
Here, for an arbitrary tensor $A^{\mu\nu}$, we define the trace-less symmetric tensor $A^{<\mu\nu>} = \frac{1}{2}[(\Delta^{\mu\alpha}\Delta^{\nu\beta}+\Delta^{\nu\alpha}\Delta^{\mu\beta})-\frac{2}{3}\Delta^{\mu\nu}\Delta^{\alpha\beta}]A_{\alpha \beta}$.

During the simulation, we choose the specific shear viscosity $C_{\eta_v}$ and baryon diffusion coefficient $\kappa_B$ as free parameters,
which are related to $\eta_v$ and $C_B$ as follows:
\begin{align}
C_{\eta_v} &= \frac{\eta_v T}{e+P}, \label{eq:C_shear}\\
\kappa_B &= \frac{C_B}{T}n\left[\frac{1}{3} \cot \left(\frac{\mu_B}{T}\right)-\frac{nT}{e+P}\right] \,, \label{eq:CB}
\end{align}
where $\mu_B$ stands for the baryon chemical potential.
The relaxation times are chosen as
\begin{equation}
    \tau_{\pi} = \frac{5C_{\eta_v}}{T},\;\;\;\tau_V = \frac{C_B}{T}.
\end{equation}
The NEOS-BQS equation of state (EOS) \citep{Monnai:2019hkn,Monnai:2021kgu} is supplied to close the equations of motion for relativistic hydrodynamics.
The NEOS-BQS EOS connects the lattice QCD and hadron gas EOS with a smooth crossover under the strangeness neutrality and electric charge density $n_Q = 0.4n_B$ condition.

Before end this subsection, we list all other parameters used in the initial conditions and hydrodynamics evolution in Table \ref{table:amptparameter}. And the $C_B$ defined in Eq. (\ref{eq:CB}) is a free parameter in general. We will discuss the $C_B$ dependence in Sec. \ref{subsec:baryon}.

\subsection{Particlization} \label{subsec:particlization}
With the QGP fireball expansion, the medium will convert to soft particles according to the Cooper-Frye formula when the the local energy density of medium cool down to  0.4~GeV/fm$^3$:
\begin{align}
E\frac{dN}{d^3p} = \frac{g_i}{(2\pi)^3}\int_{\Sigma} p^{\mu}d\Sigma_{\mu}(f_{eq}+\delta f_{\pi}+\delta f_{V})\,.
\end{align}
Here $g_i$ is the degeneracy for identified hadrons; $d\Sigma_{\mu}$ is the hyper-surface element which is determined from the projection method \cite{Pang:2012he}; $f_{\rm eq}$, $\delta f_{\pi}$ and $\delta f_{V}$ are thermal equilibrium distribution and out-of-equilibrium corrections, which take the following forms \citep{McNelis:2021acu,McNelis:2019auj,Denicol:2018wdp}
\begin{align}
	f_{\rm eq}(x,p) &= \frac{1}{\exp \left[(p_{\mu}U^{\mu} - B\mu_B \right)/T_f] \pm 1} \, ,\\
	\delta f_{\pi}(x,p) &= f_{\rm eq}(1\pm f_{\rm eq}) \frac{p_{\mu}p_{\nu}\pi^{\mu\nu}}{2T^2_f(e+P)}, \\
	\delta f_V(x,p) &= f_{\rm eq}(1\pm f^{\rm eq})\left(\frac{n}{e+P}-\frac{B}{U^{\mu}p_{\mu}}\right)\frac{p^{\mu}V_{\mu}}{\kappa_B/ \tau_V } ,
\end{align}
where $T_f$ is the chemical freeze-out temperature,
$B$ is the baryon number for the identified baryon. After the particlization of the fluid, we simulate the hadrons by SMASH model \citep{Weil:2016zrk,Schafer:2019edr,Mohs:2019iee,Hammelmann:2019vwd,Mohs:2020awg} again.

\subsection{Spin polarization}\label{subsec:SpinPolarization}

In non-central heavy-ion collisions, the quarks are polarized due to the huge initial orbital angular momentum of the QGP fireball.
One can assume that the quarks reach to the local (thermal) equilibrium at freezeout hyper-surface.
As a common strategy in the community, we further assume that the spins of quarks or hadrons are not modified during the pariclization and hadronic cascade.
The polarization pseudo vector for spin-$\frac{1}{2}$ particles can be evaluated by the modified Cooper-Frye formula \citep{Becattini:2013fla,Fang:2016vpj},
\begin{equation}
\mathcal{S}^{\mu}(\mathbf{p})=\frac{\int d \Sigma \cdot p \mathcal{J}_{5}^{\mu}(p, X)}{2 m \int d \Sigma \cdot \mathcal{N}(p, X)}
\end{equation}
where $\mathcal{J}^{\mu}_5$ and $\mathcal{N}^{\mu}(p, X)$ are axial-charge current density and the number density of fermions in phase space, respectively.
Following the results from quantum kinetic theory \citep{Yi:2021ryh,Hidaka:2017auj,Yi:2021unq},  $\mathcal{S}^{\mu}(\mathbf{p})$ can be decomposed into different sources,
\begin{eqnarray}
\mathcal{S}^{\mu}(\mathbf{p}) & = & \mathcal{S}_{\textrm{thermal}}^{\mu}(\mathbf{p})+\mathcal{S}_{\textrm{shear}}^{\mu}(\mathbf{p})+\mathcal{S}_{\textrm{accT}}^{\mu}(\mathbf{p})   \nonumber \\
& &+\mathcal{S}_{\textrm{chemical}}^{\mu}(\mathbf{p})+\mathcal{S}_{\textrm{EB}}^{\mu}(\mathbf{p}),
\end{eqnarray}
where
\begin{eqnarray}
\mathcal{S}_{\textrm{thermal}}^{\mu}(\mathbf{p}) & = & \int d\Sigma^{\sigma}F_{\sigma}\epsilon^{\mu\nu\alpha\beta}p_{\nu}\partial_{\alpha}\frac{u_{\beta}}{T},\nonumber \\
\mathcal{S}_{\textrm{shear}}^{\mu}(\mathbf{p}) & = & \int d\Sigma^{\sigma}F_{\sigma}  \frac{\epsilon^{\mu\nu\alpha\beta}p_{\nu} u_{\beta}}{(u\cdot p)T}
 \nonumber \\
 & &\times  p^{\rho}(\partial_{\rho}u_{\alpha}+\partial_{\alpha}u_{\rho}-u_{\rho}Du_{\alpha}) \nonumber \\
\mathcal{S}_{\textrm{accT}}^{\mu}(\mathbf{p}) & = & -\int d\Sigma^{\sigma}F_{\sigma}\frac{\epsilon^{\mu\nu\alpha\beta}p_{\nu}u_{\alpha}}{T}
\left(Du_{\beta}-\frac{\partial_{\beta}T}{T}\right),\nonumber \\
\mathcal{S}_{\textrm{chemical}}^{\mu}(\mathbf{p}) & = & 2\int d\Sigma^{\sigma}F_{\sigma}\frac{1}{(u\cdot p)}\epsilon^{\mu\nu\alpha\beta}p_{\alpha}u_{\beta}\partial_{\nu}\frac{\mu}{T},\nonumber \\
\mathcal{S}_{\textrm{EB}}^{\mu}(\mathbf{p}) & = & 2\int d\Sigma^{\sigma}F_{\sigma}\left[\frac{\epsilon^{\mu\nu\alpha\beta}p_{\alpha}u_{\beta}E_{\nu}}{(u\cdot p)T}+\frac{B^{\mu}}{T}\right],\nonumber \label{eq:S_all}  \\
&
\end{eqnarray}
with
\begin{eqnarray}
F^{\mu} &=& \frac{\hbar}{8m_{\Lambda}\Phi(\mathbf{p})}p^{\mu}f_{eq}(1-f_{eq}), \nonumber \\
\Phi(\mathbf{p}) &=& \int d\Sigma^{\mu}p_{\mu}f_{eq}.
\label{eq:def_N}
\end{eqnarray}
The above equations represent the polarization induced by the thermal vorticity, the shear tensor,
the fluid acceleration minus temperature gradient (accT), the gradient of chemical potential over temperature,
and the external electromagnetic fields, respectively.
Since the electromagnetic fields decay rapidly, we neglect the $\mathcal{S}_{\textrm{EB}}^{\mu}(\mathbf{p})$ in our simulations.
$S^\mu_{\textrm{shear}}$ and $S^\mu_{\textrm{chemical}}$ are also named shear-induced polarization (SIP) and baryonic spin Hall effect (SHE), respectively.
In general, the above expression for the polarization can also be derived
from different models, e.g. the statistic model \cite{Becattini:2021suc,Becattini:2021iol}
and Kubo formula \cite{Liu:2020dxg,Liu:2021uhn,Fu:2021pok,Fu:2022myl}.

In the experiment, the polarization of $\Lambda$ and $\overline{\Lambda}$ are measured in their own rest frames. Therefore, we express the polarization pseudo vector in the rest frame of $\Lambda$ and $\overline{\Lambda}$, named $\vec{P}^{*}(\mathbf{p})$,
by taking the Lorenz transformation,
\begin{eqnarray}
\vec{P}^{*}(\mathbf{p}) = \vec{P}(\mathbf{p})-\frac{\vec{P}(\mathbf{p}) \cdot \vec{p}}{p^{0}(p^{0}+m)}\vec{p},
\end{eqnarray}
where \begin{eqnarray}
P^{\mu}(\mathbf{p}) \equiv \frac{1}{s} \mathcal{S}^{\mu}(\mathbf{p}),
\end{eqnarray}
with $s=1/2$ being the spin of the particle.
Finally, the local polarization is given by the averaging over momentum and rapidity as follows,
\begin{eqnarray}
\langle \vec{P}(\phi_p) \rangle = \frac{\int_{y_{\text{min}}}^{y_{\text{max}}}dy \int_{p_{T\text{min}}}^{p_{T\text{max}}}p_Tdp_T
[ \Phi (\mathbf{p})\vec{P}^{*}(\mathbf{p})]}{\int_{y_{\text{min}}}^{y_{\text{max}}}dy \int_{p_{T\text{min}}}^{p_{T\text{max}}}p_Tdp_T \Phi(\mathbf{p}) },
\end{eqnarray}
where $\phi_p$ is the azimuthal angle.
For convenience, we follow Eqs. (\ref{eq:S_all}) and use the letters in the lower indices for the polarization induced by different sources, e.g. $P^z_{\textrm{shear}}$ stands for the local polarization along the beam direction induced by the shear tensor, and the $P^i(\phi_p)$ for the total local polarization.

In the current work, we focus on polarization at $20-50\%$ centrality
in $\sqrt{s_{NN}}=7.7,\; 27,\; 62.4$ GeV Au+Au collisions and
take the mass of $\Lambda$ or $\overline{\Lambda}$ as $m = 1.116$ GeV. The region of transverse momentum and rapidity is chosen as $p_T \in [0.5,3.0]$ and $y\in [-1,1]$.

\section{Numerical Results}
\label{results}

\begin{figure*}[thb]
\includegraphics[scale=0.285]{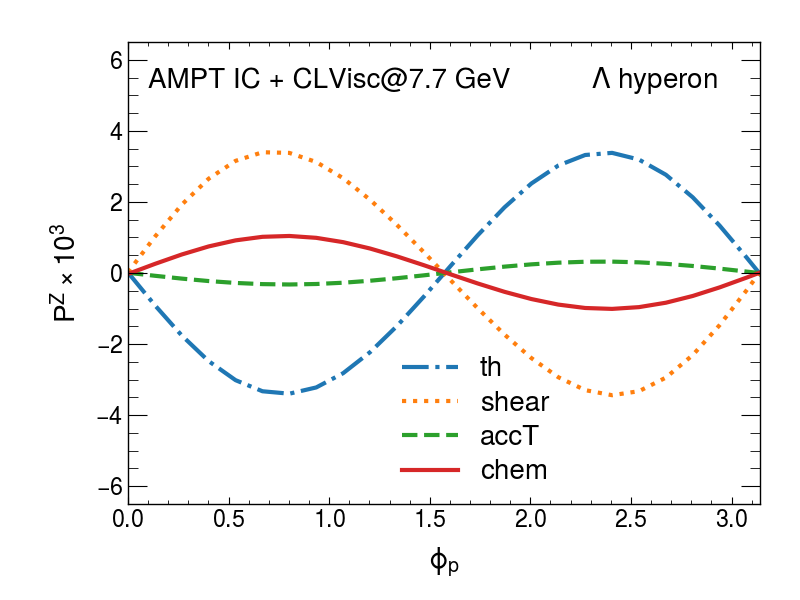}
\includegraphics[scale=0.285]{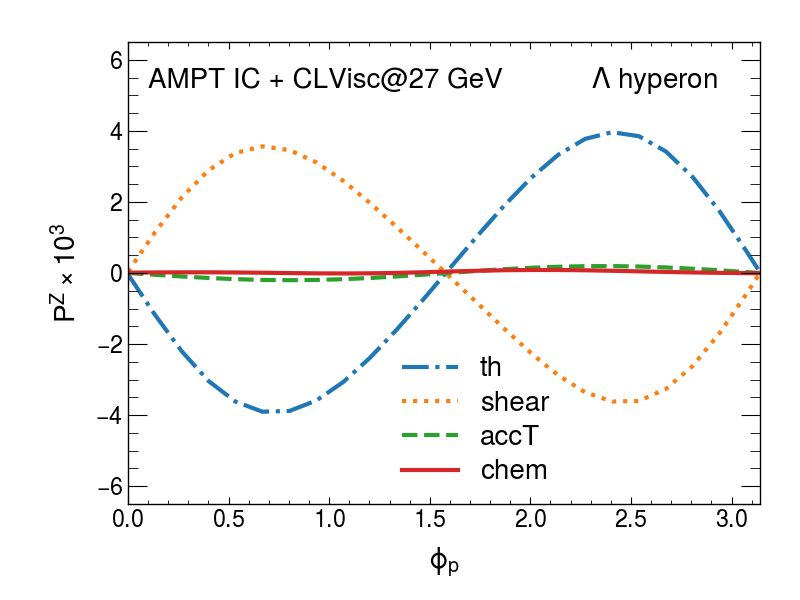}
\includegraphics[scale=0.285]{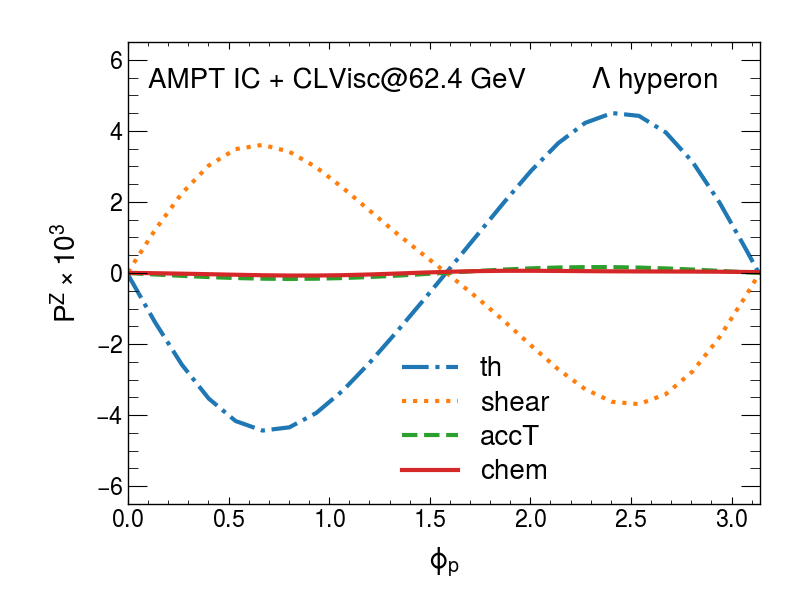}
\includegraphics[scale=0.285]{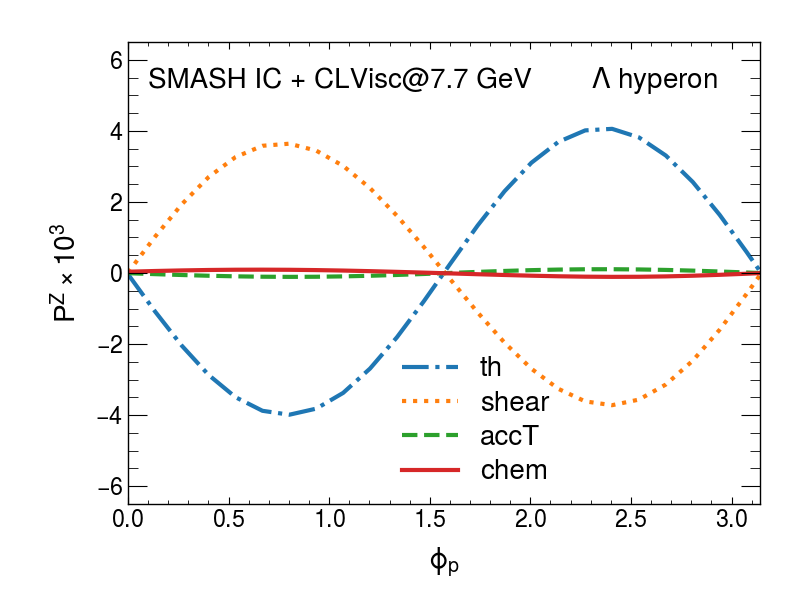}
\includegraphics[scale=0.285]{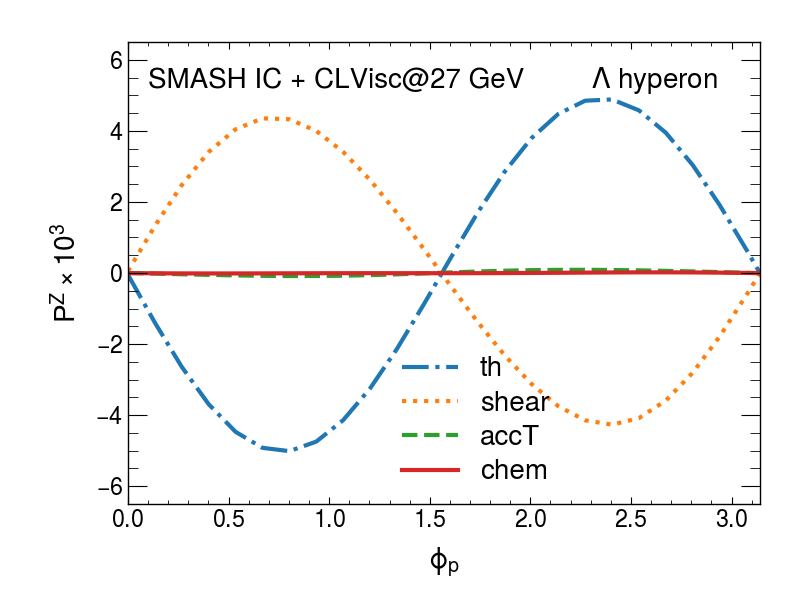}
\includegraphics[scale=0.285]{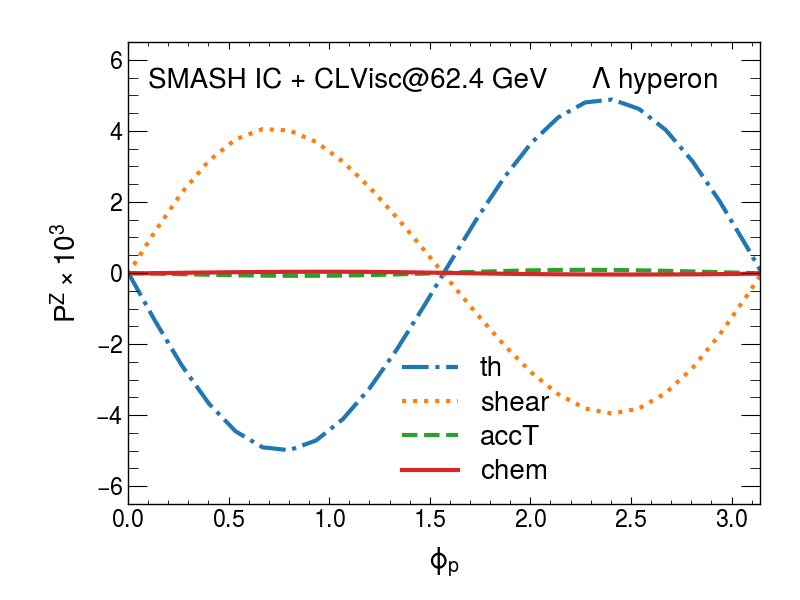}
\caption{The local polarization along the beam direction of $\Lambda$ hyperons, $P^z$, as a function of azimuthal angle $\phi_p$ in $20-50\%$ centrality  at $\sqrt{s_{NN}}=7.7, \;27,\;62.4 $GeV Au+Au collisions with AMPT and SMASH initial conditions. The results are set up with $p_T \in [0.5,3.0]$ and $y\in [-1,1]$. The coefficient $C_B$ in Eq.~(\ref{eq:CB}) is set to be zero. The different colors stand for the separated contribution induced by the thermal vorticity, SIP, SHE and the acceleration terms.}
\label{fig:pz_mode}
\includegraphics[scale=0.285]{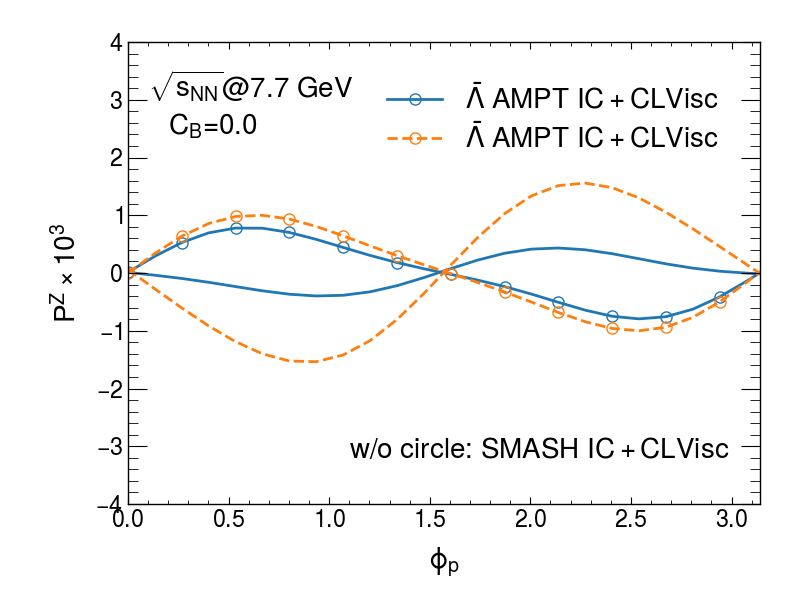}
\includegraphics[scale=0.285]{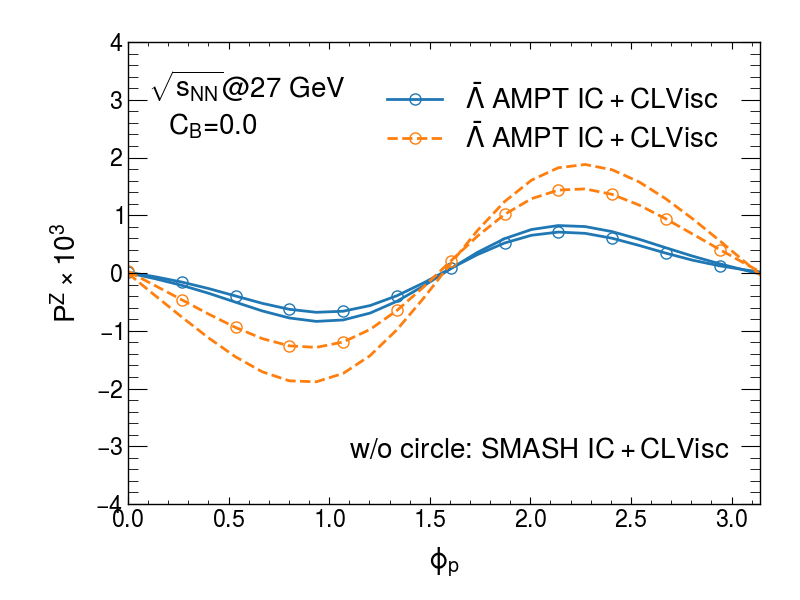}
\includegraphics[scale=0.285]{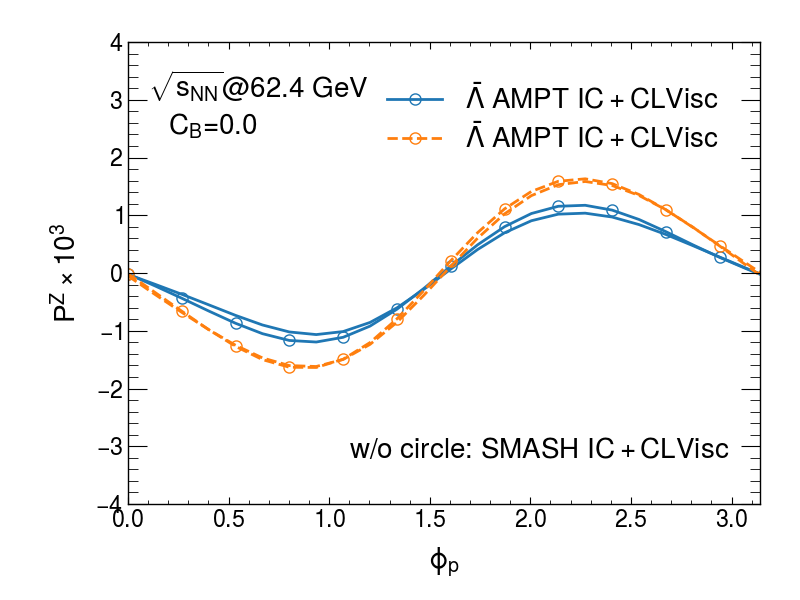}
\caption{
The total local polarization along the beam direction of $\Lambda$ and $\overline{\Lambda}$ hyperons, $P^z$, as a function of azimuthal angle $\phi_p$ in $20-50\%$ centrality  at $\sqrt{s_{NN}}=7.7, \;27,\;62.4 $GeV Au+Au collisions with AMPT and SMASH initial conditions. The results are set up with $p_T \in [0.5,3.0]$, $y\in [-1,1]$ and $C_B=0$ in Eq.~(\ref{eq:CB}).
The line with or without circle stand for the AMPT or SMASH initial condition. The blue solid and orange dashed lines denote the results for $\Lambda$ and $\overline{\Lambda}$ hyperons.
}
\label{fig:pz_polar_total}
\end{figure*}

\begin{figure*}[thb]
\includegraphics[scale=0.285]{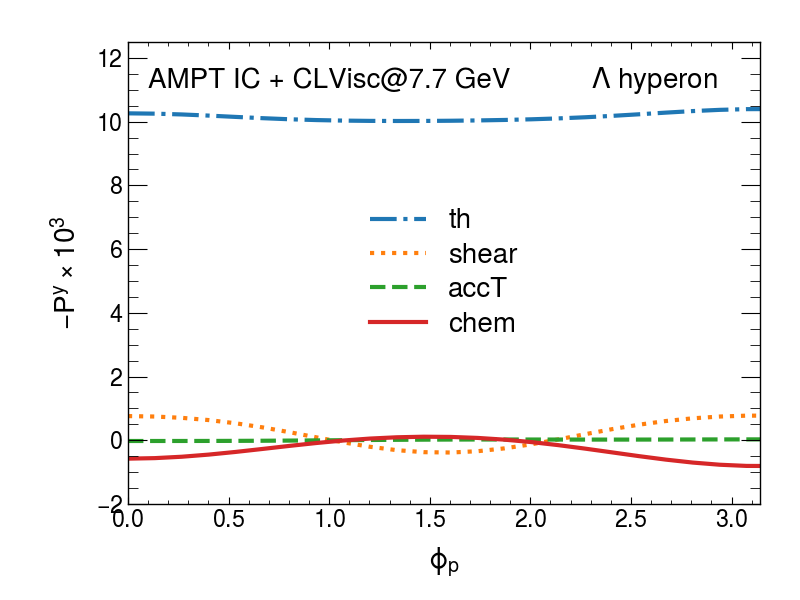}
\includegraphics[scale=0.285]{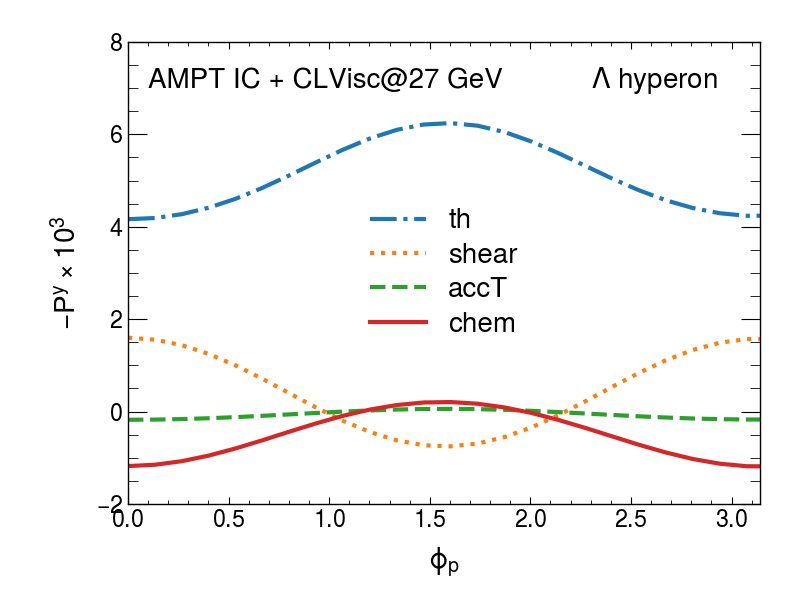}
\includegraphics[scale=0.285]{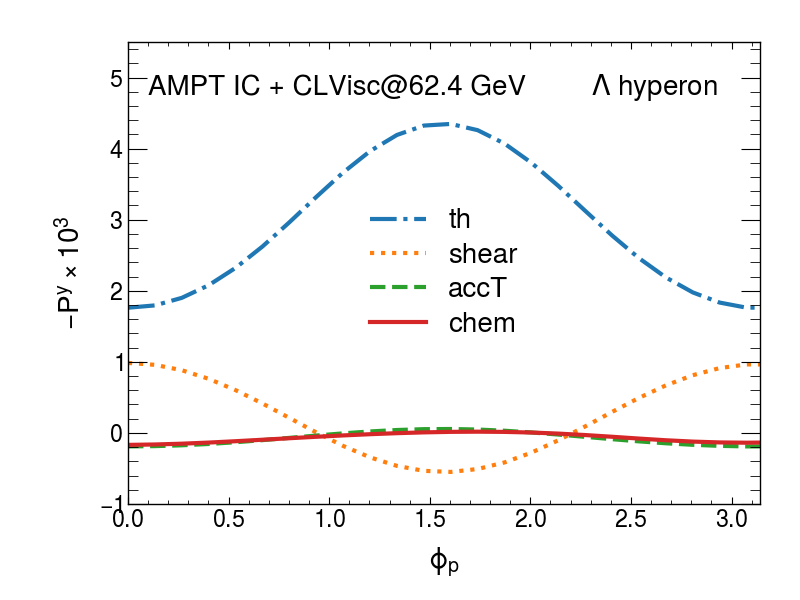}
\includegraphics[scale=0.285]{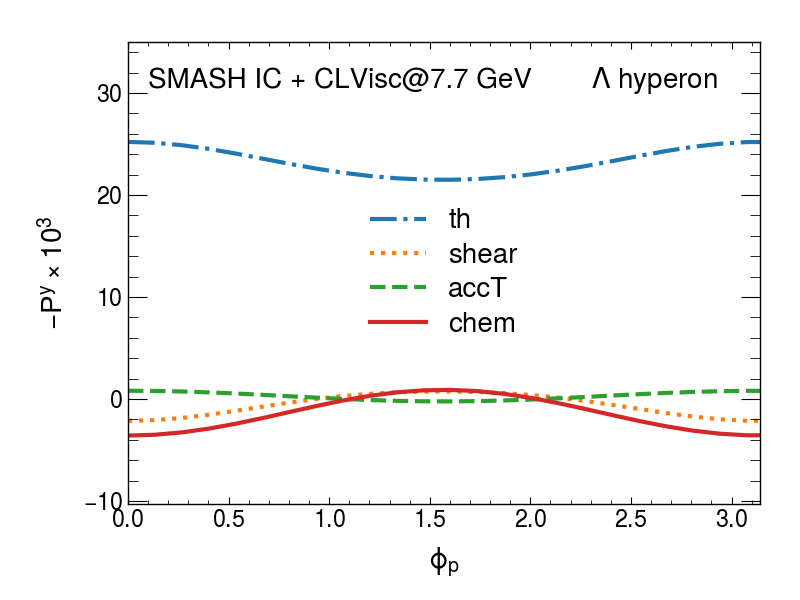}
\includegraphics[scale=0.285]{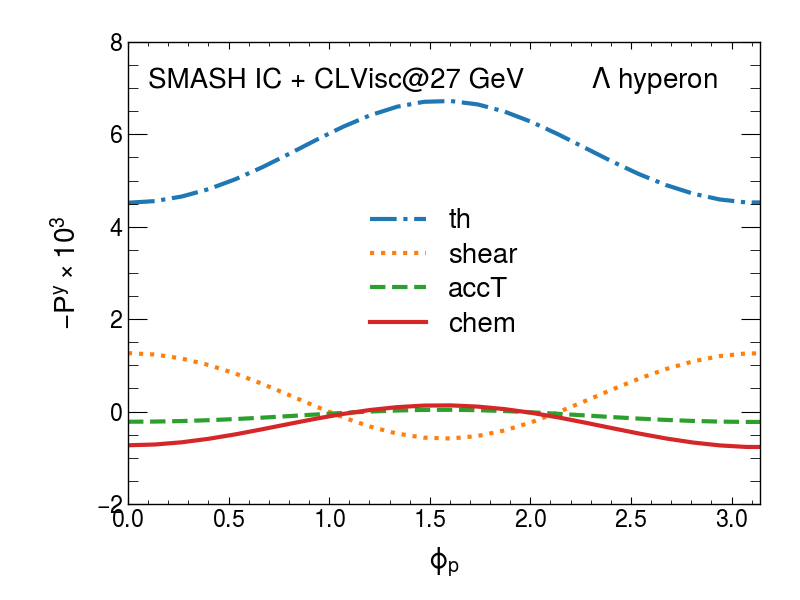}
\includegraphics[scale=0.285]{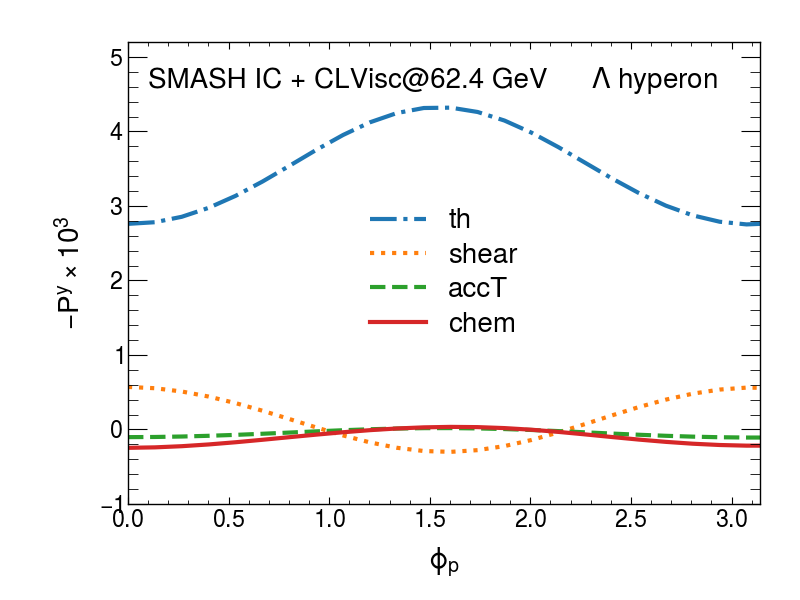}
\caption{The local polarization along the $y$ direction of $\Lambda$ hyperons, $P^y$, as a function of azimuthal angle $\phi_p$ in $20-50\%$ centrality  at $\sqrt{s_{NN}}=7.7, \;27,\;62.4 $GeV Au+Au collisions with AMPT and SMASH initial conditions. The results are set up with $p_T \in [0.5,3.0]$ and $y\in [-1,1]$.The coefficient $C_B$ in Eq.~(\ref{eq:CB}) is set to be zero. The different colors stand for the separated contribution induced by the thermal vorticity, SIP, SHE and the acceleration terms.
}
\label{fig:py_mode}
\includegraphics[scale=0.285]{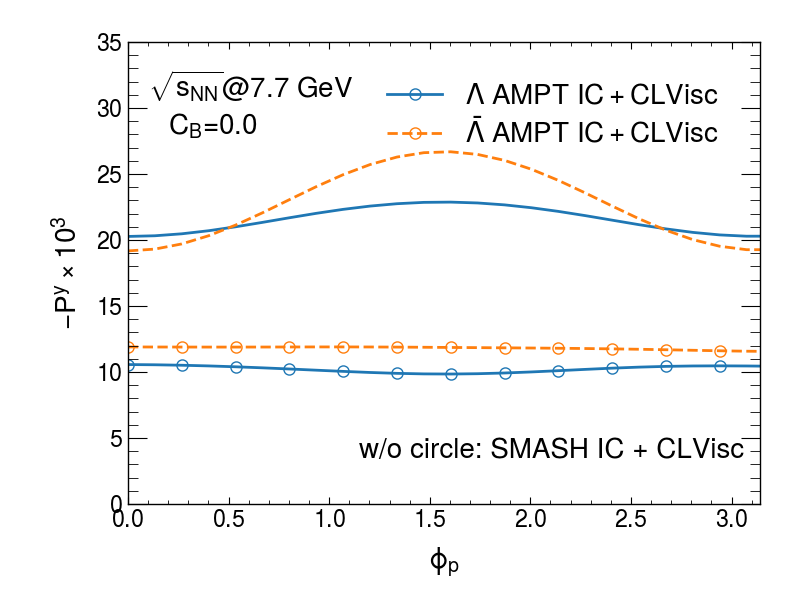}
\includegraphics[scale=0.285]{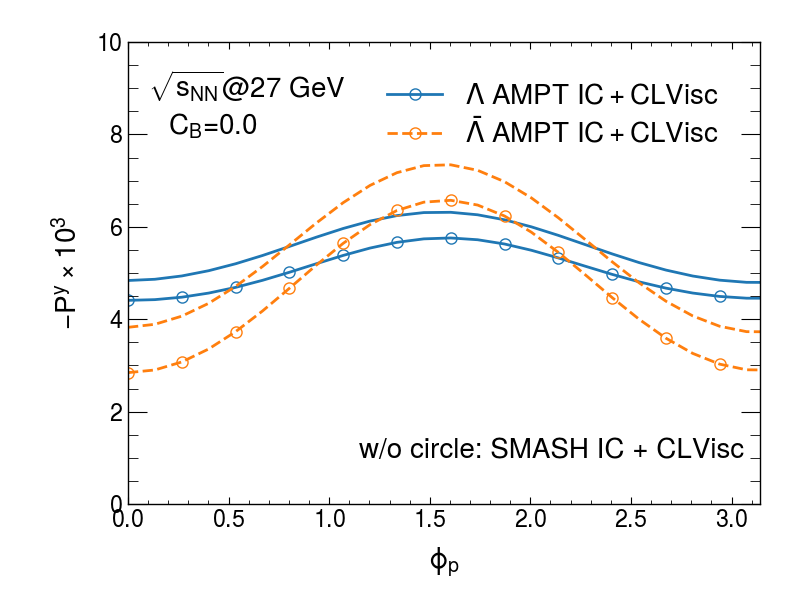}
\includegraphics[scale=0.285]{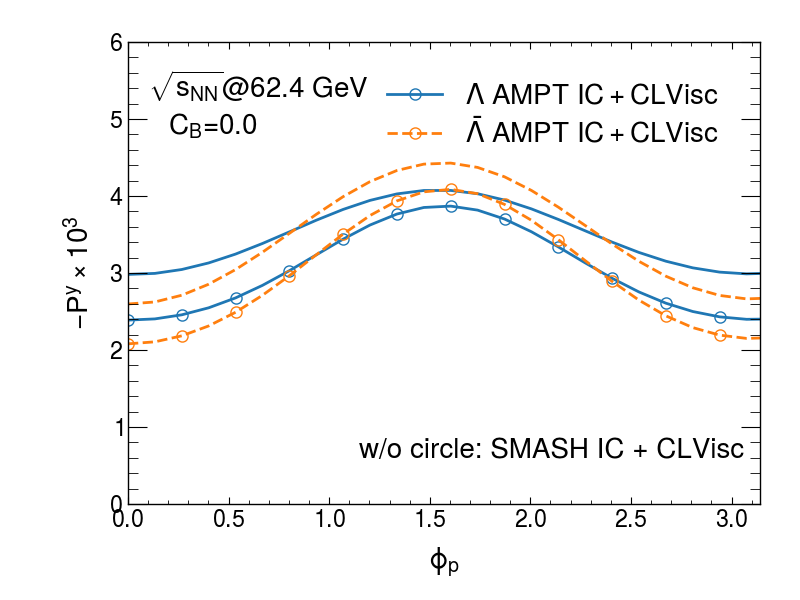}
\caption{
The total local polarization along the $y$ direction of $\Lambda$ and $\overline{\Lambda}$ hyperons, $P^y$, as a function of azimuthal angle $\phi_p$ in $20-50\%$ centrality  at $\sqrt{s_{NN}}=7.7, \;27,\;62.4 $GeV Au+Au collisions with AMPT and SMASH initial conditions. The results are set up with $p_T \in [0.5,3.0]$, $y\in [-1,1]$ and $C_B=0$ in Eq.~(\ref{eq:CB}).
The line with or without circle stand for the AMPT or SMASH initial condition. The blue solid and orange dashed lines denote the results for $\Lambda$ and $\overline{\Lambda}$ hyperons.
}
\label{fig:py_mode_total}
\end{figure*}

\begin{figure*}[thb]
\includegraphics[scale=0.285]{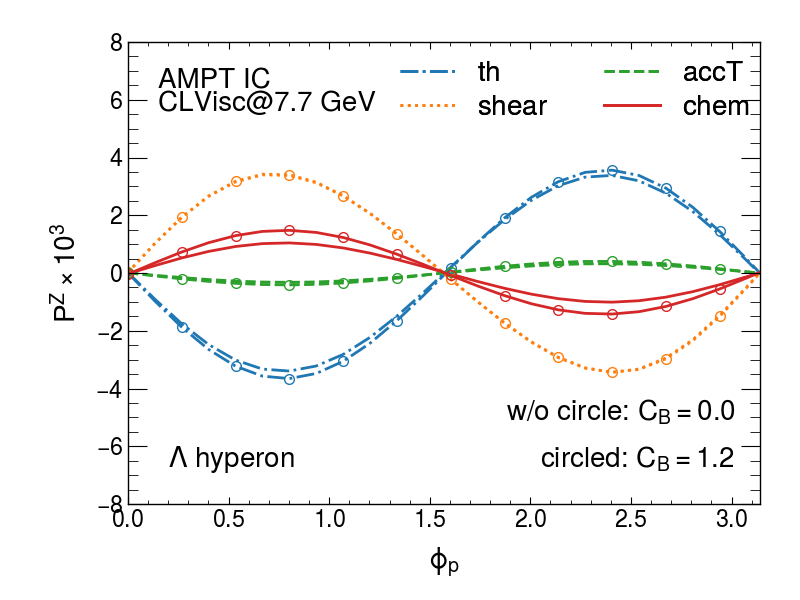}
\includegraphics[scale=0.285]{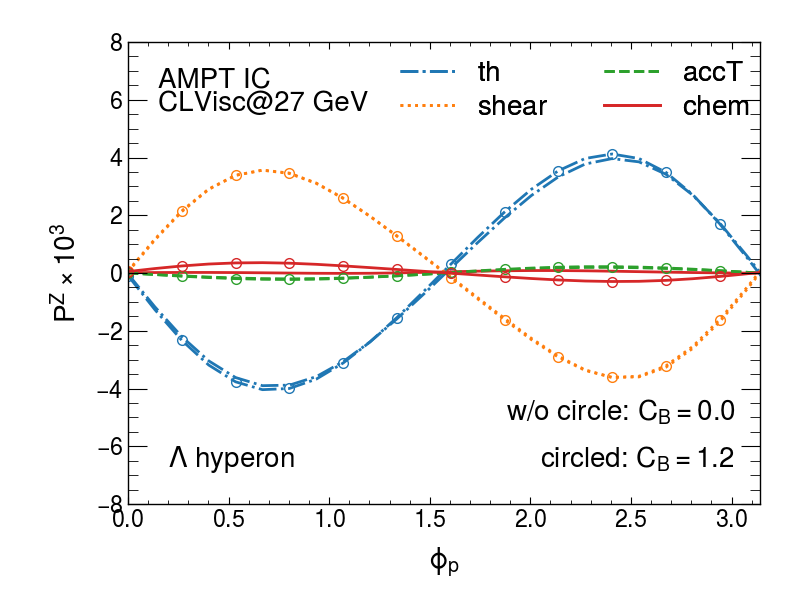}
\includegraphics[scale=0.285]{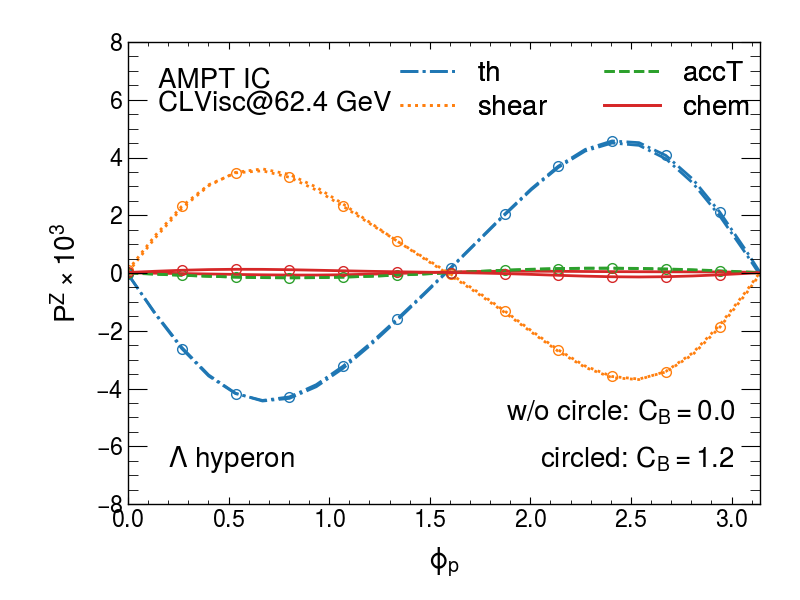}
\includegraphics[scale=0.285]{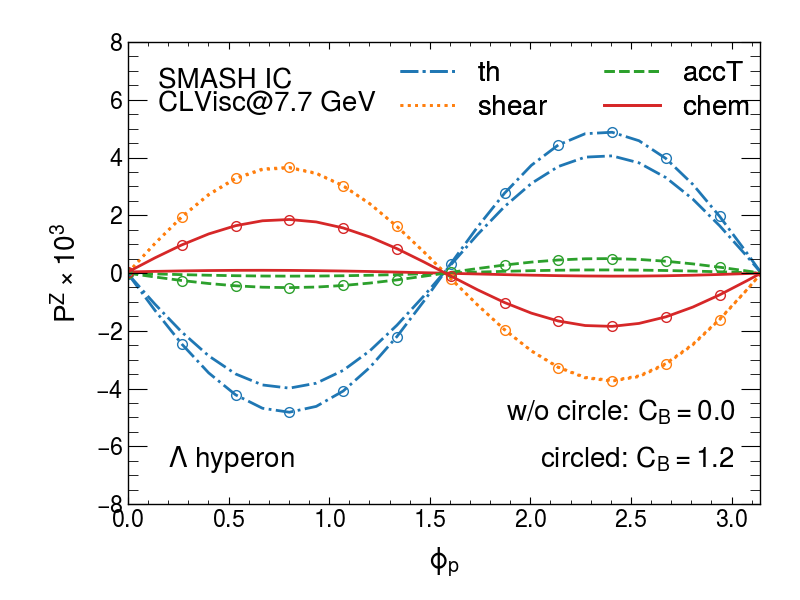}
\includegraphics[scale=0.285]{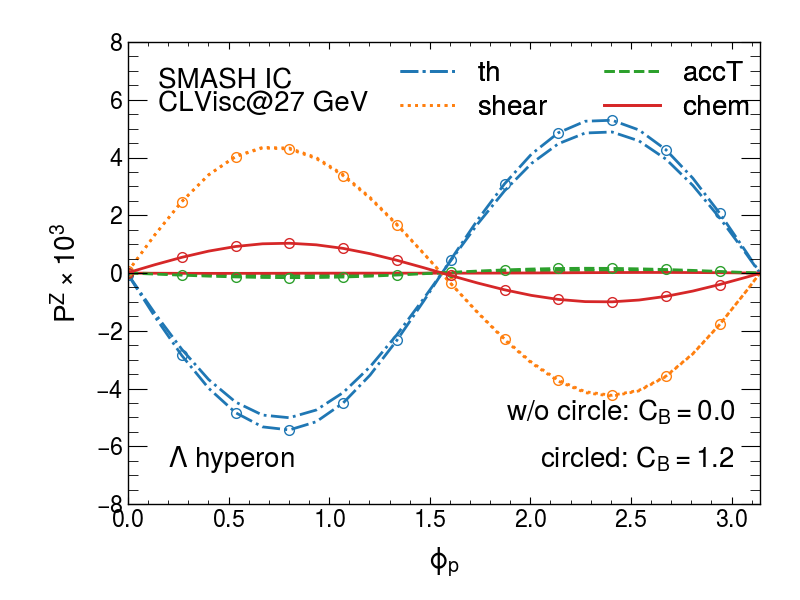}
\includegraphics[scale=0.285]{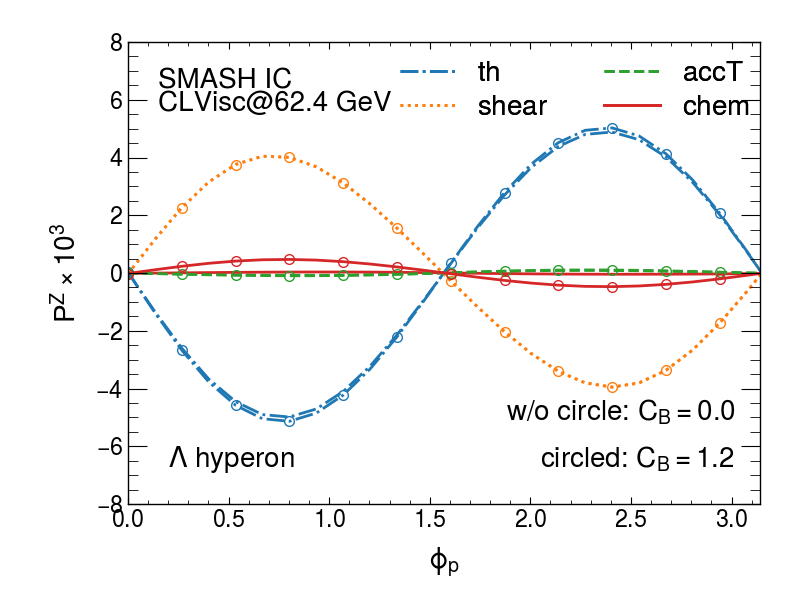}
\caption{The local polarization along the beam direction of $\Lambda$ hyperons, $P^z$, as a function of azimuthal angle $\phi_p$ in $20-50\%$ centrality  at $\sqrt{s_{NN}}=7.7, \;27,\;62.4 $GeV Au+Au collisions with AMPT and SMASH initial conditions and different baryon diffusion coefficient $C_B$. The results are set up with $p_T \in [0.5,3.0]$ and $y\in [-1,1]$. The different colors stand for the separated contribution induced by the thermal vorticity, SIP, SHE and the acceleration terms. The line without or with circle denotes $C_B =0$ and $C_B=1.2$, respectively. }
\label{fig:pz_mode_cb}
\includegraphics[scale=0.285]{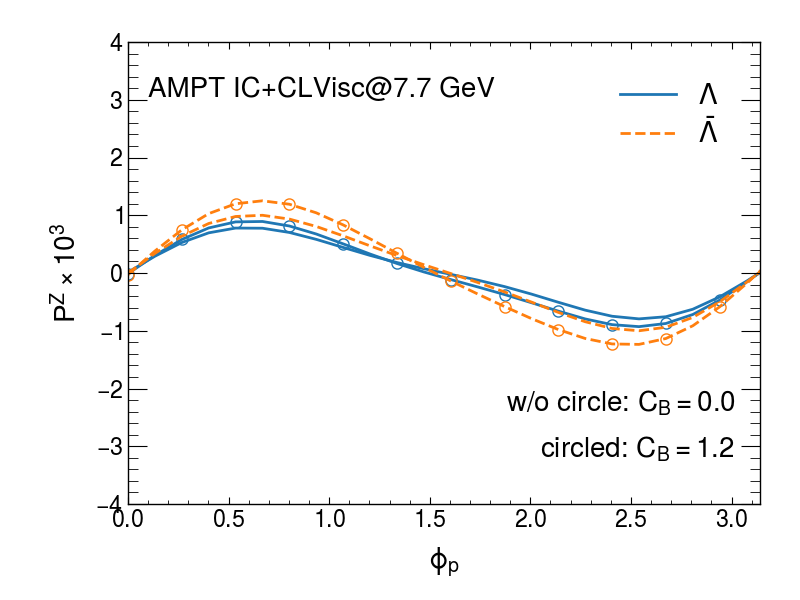}
\includegraphics[scale=0.285]{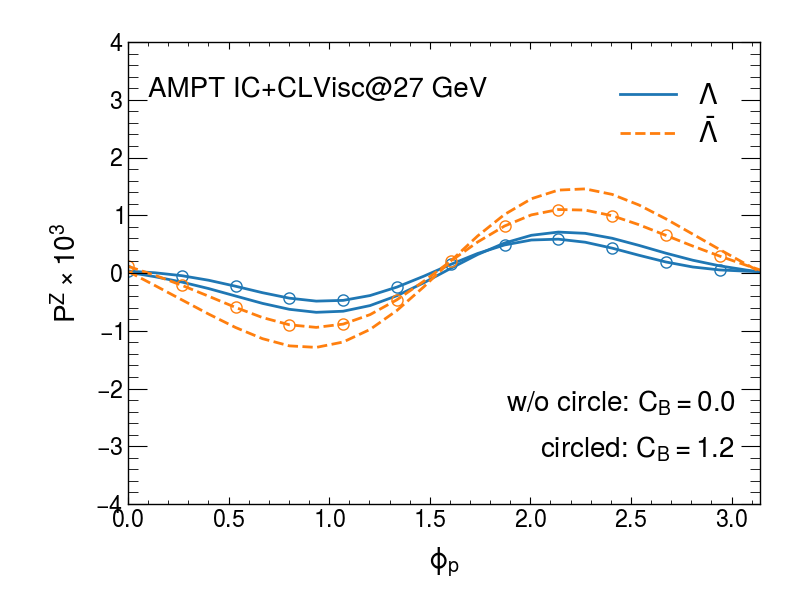}
\includegraphics[scale=0.285]{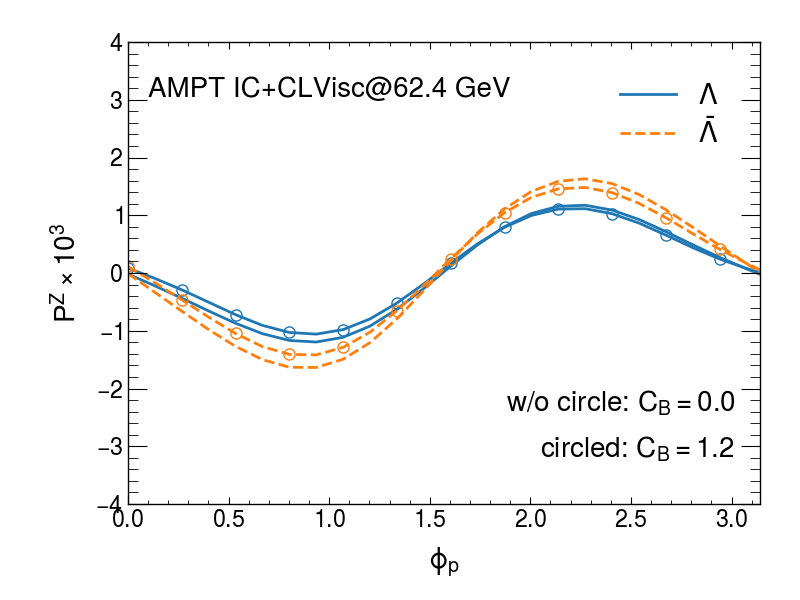}
\includegraphics[scale=0.285]{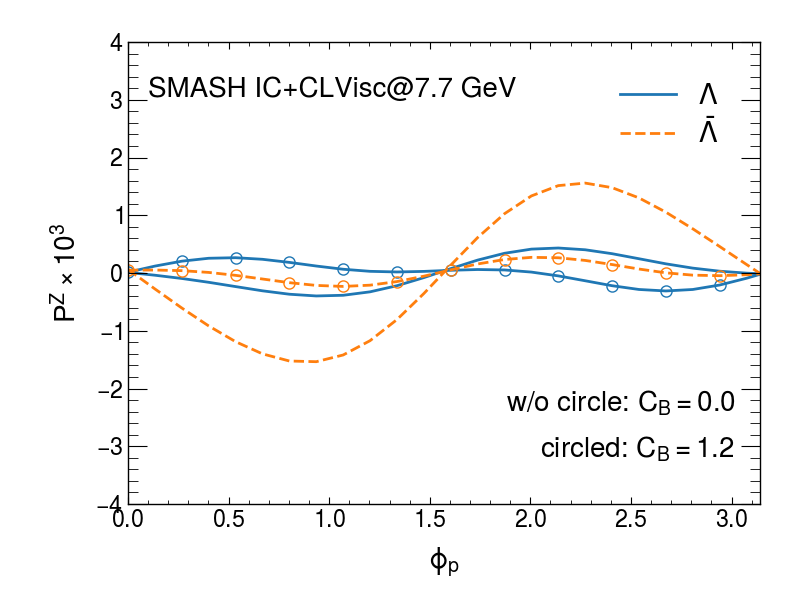}
\includegraphics[scale=0.285]{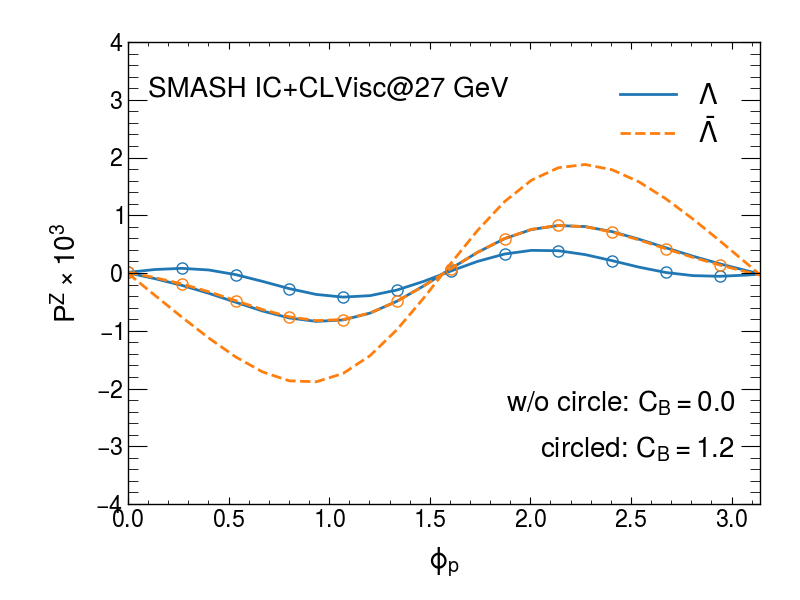}
\includegraphics[scale=0.285]{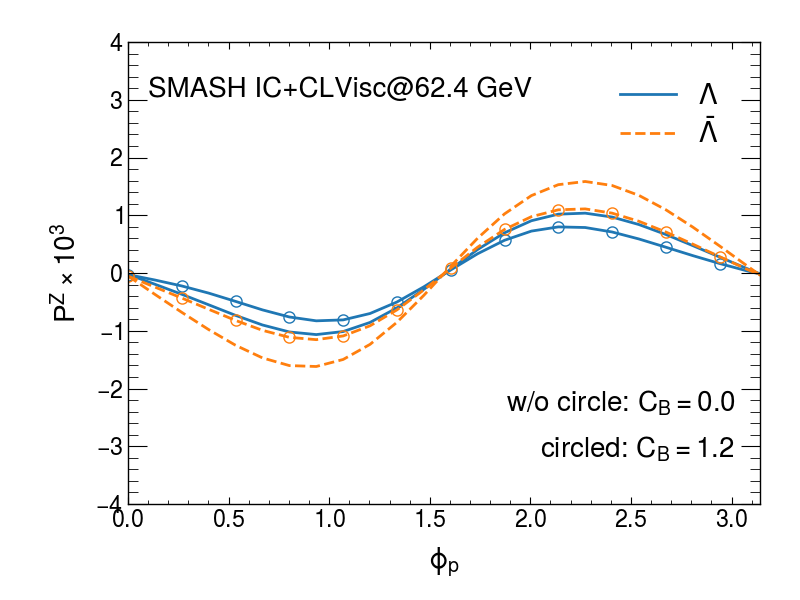}
\caption{
The total local polarization along the beam direction of $\Lambda$ and $\overline{\Lambda}$ hyperons, $P^z$, as a function of azimuthal angle $\phi_p$ in $20-50\%$ centrality  at $\sqrt{s_{NN}}=7.7, \;27,\;62.4 $GeV Au+Au collisions with AMPT and SMASH initial conditions and different baryon diffusion coefficient $C_B$. The results are set up with $p_T \in [0.5,3.0]$ and $y\in [-1,1]$.
The line without or with circle denotes $C_B =0$ and $C_B=1.2$, respectively.
}
\label{fig:pz_total_cb}
\end{figure*}

\begin{figure*}[thb]
\includegraphics[scale=0.285]{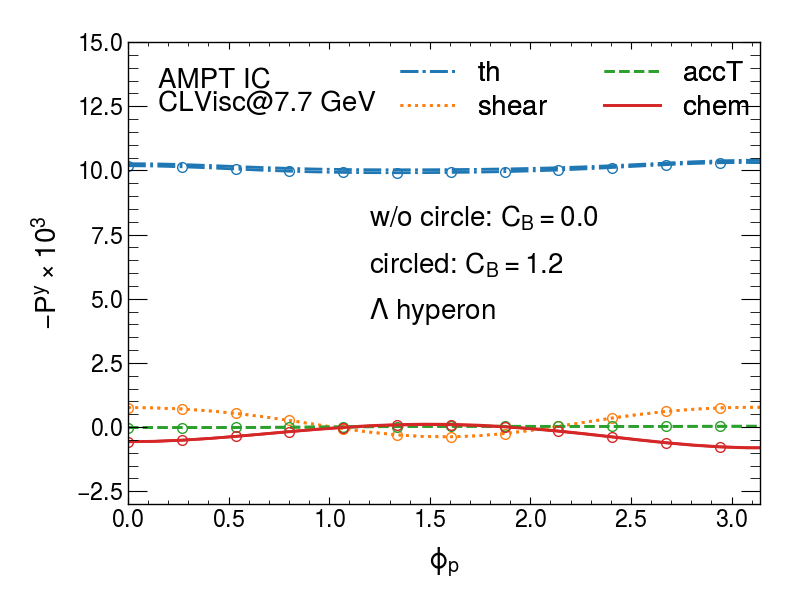}
\includegraphics[scale=0.285]{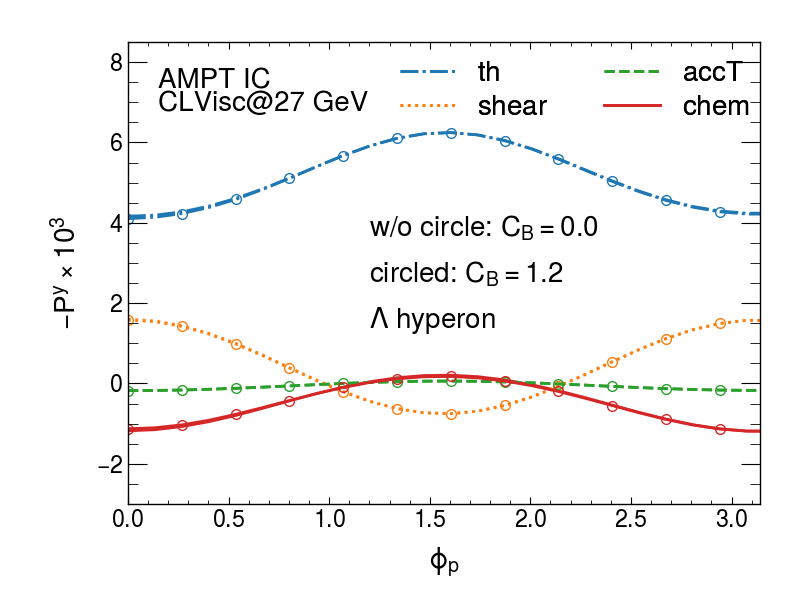}
\includegraphics[scale=0.285]{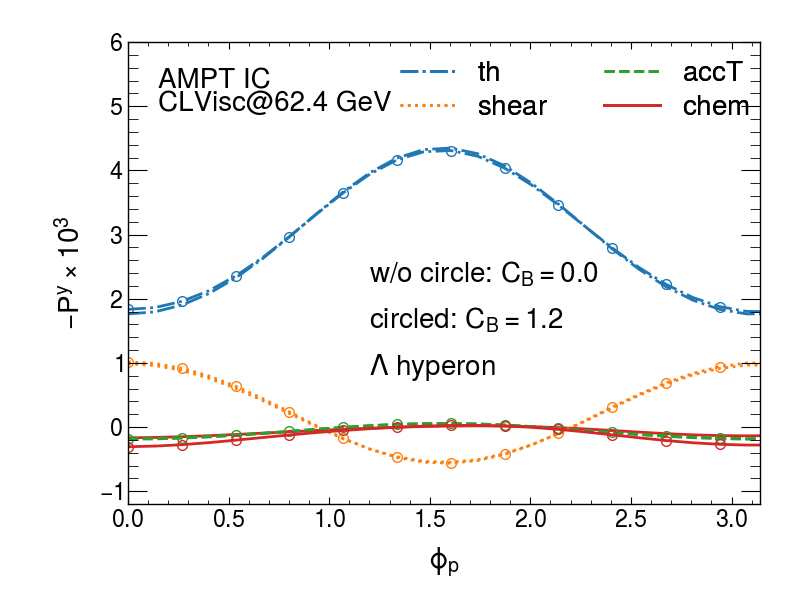}
\includegraphics[scale=0.285]{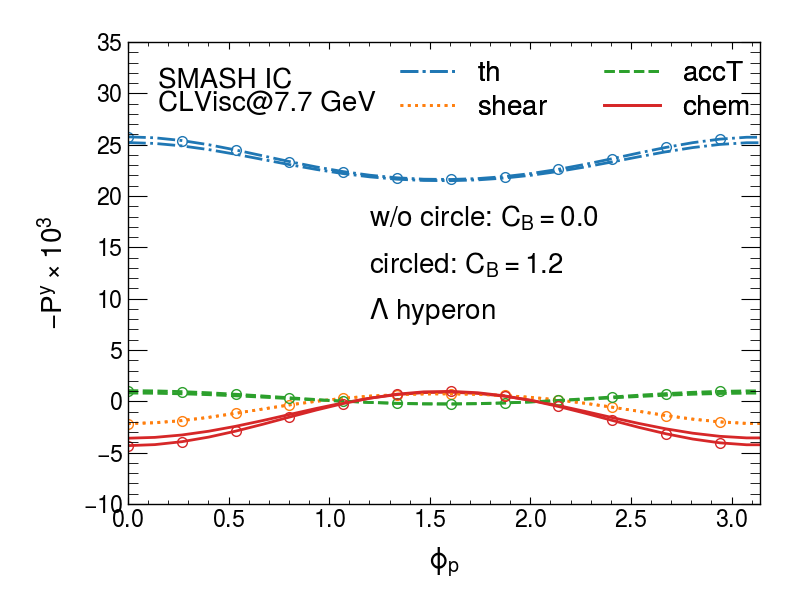}
\includegraphics[scale=0.285]{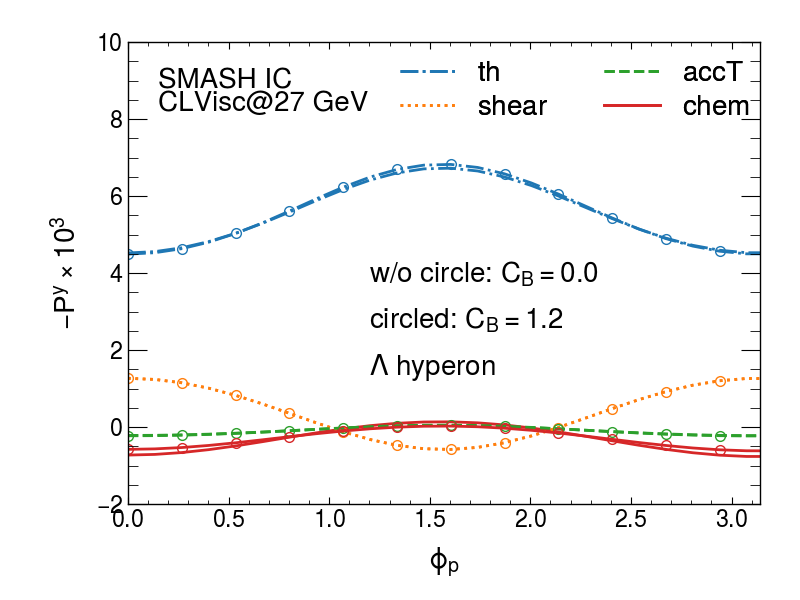}
\includegraphics[scale=0.285]{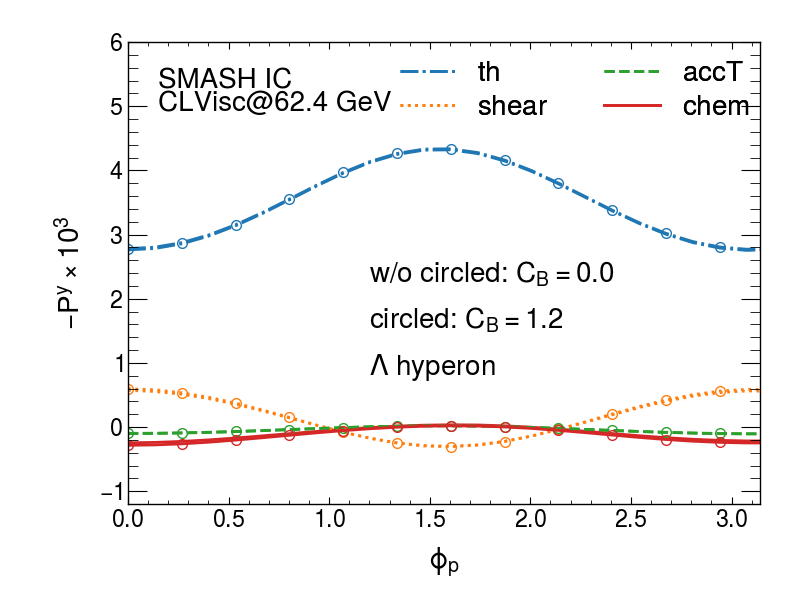}
\caption{
The local polarization along $y$ direction of $\Lambda$ hyperons, $P^y$, as a function of azimuthal angle $\phi_p$ in $20-50\%$ centrality  at $\sqrt{s_{NN}}=7.7, \;27,\;62.4 $GeV Au+Au collisions with AMPT and SMASH initial conditions and different baryon diffusion coefficient $C_B$. The results are set up with $p_T \in [0.5,3.0]$ and $y\in [-1,1]$. The different colors stand for the separated contribution induced by the thermal vorticity, SIP, SHE and the acceleration terms. The line without or with circle denotes $C_B =0$ and $C_B=1.2$, respectively.
}
\label{fig:py_mode_cb}
\end{figure*}

In this section, we present the numerical results for the spin polarization in Au+Au collisions at RHIC-BES energies
using the (3+1) dimensional CLVisc hydrodynamics framework with AMPT and SMASH initial conditions.
We first present the results for the local polarization of $\Lambda$ and $\overline{\Lambda}$ hyperons induced by different sources at different collision energies in Sec. \ref{subsec:local}.
We discuss the contributions from the SHE and compare the results from AMPT and SMASH initial conditions.
Then we study the baryon diffusion dependence of the polarization in Sec. \ref{subsec:baryon}.
At last, we also show the global polarization of $\Lambda$ and $\overline{\Lambda}$ hyperons.

\subsection{Local polarization and SHE} \label{subsec:local}

In  Fig.~\ref{fig:pz_mode}, we plot the local polarization of $\Lambda$ hyperons along beam direction, $P^z$, contributed from different components
using AMPT and SMASH initial conditions at $20-50\%$ centrality in $\sqrt{s_{NN}} = 7.7, \; 27, \;62.4 $GeV Au+Au collisions.
For both AMPT and SMASH initial conditions, the polarization induced by SHE ($P^z_{\rm chem}$) and SIP ($P^z_{\rm shear}$) provide the sine contribution to longitudinal polarization $P^z$,
while the polarization from thermal vorticity $P^z_{\rm th}$ and fluid acceleration $P^z_{\rm accT}$  give the opposite contribution. These results are similar to previous studies \citep{Fu:2021pok,Becattini:2021iol, Yi:2021ryh, Fu:2022myl}.

Let us focus on the collision energy and initial condition dependences.
For the simulations with AMPT initial conditions, we find that
the longitudinal polarization induced by the thermal vorticity $P^z_{\rm th}$, shear tensor $P^z_{\rm shear}$ and the fluid acceleration $P^z_{\rm accT}$ show a weak dependence on the collision energy.
The magnitude of $P^z_{\rm accT}$ or $P^z_{\rm th}, P^z_{\rm shear}$ increases or decreases when the collision energies decrease, respectively.
In contrast, for AMPT initial condition, the longitudinal polarization induced by SHE strongly depends on the collision energy and $P^z_{\rm chem}$ grows when the collision energy decreases.

Next, we compare the results from AMPT model with the one from SMASH model.
We find that the magnitude of $P^z_{\rm th}$ and $P^z_{\rm shear}$ derived by the simulations with SMASH model are at the same order as the one with AMPT initial condition. This is expected because
the radial flow in the final state given by SMASH model is similar to the one from AMPT model. Again, we observe that $P^z_{\rm accT}$ is almost vanishing in the simulations with SMASH initial condition.
Interestingly, $P^z_{\rm chem}$ simulated from SMASH initial condition becomes almost negligible and has a weak dependence on the collision energies.
These results indicate the longitudinal polarization induced by SHE $P^z_{\rm chem}$ depends on initial conditions strongly.

By summing over the contributions from all the above sources in Fig.~\ref{fig:pz_mode}, we plot the total local spin polarization along beam direction $P^z$ for $\Lambda$ and $\bar{\Lambda}$ hyperons in Fig.~\ref{fig:pz_polar_total}  as a function of $\phi_p$ in $20-50\%$ centrality at $\sqrt{s_{NN}} = 7.7,\;27,\;62.4$GeV Au+Au collisions with AMPT and SMASH initial conditions.
At $\sqrt{s_{NN}} =  27,\; 62.4$GeV, the simulations with both AMPT and SMASH initial conditions give the similar total local polarization $P^z$.
Surprisingly, the total local polarization $P^z$ at $\sqrt{s_{NN}}=  7.7$GeV from AMPT initial conditions is significant different with the one from SMASH initial conditions.
SMASH initial condition gives an opposite sign for $P^z$ in contrast to AMPT initial condition due to the contributions from $P^z_{\rm chem}$. Again, such results originate from the difference of the initial baryon density profiles in two models.
Another important observation is that the longitudinal polarization for $\bar{\Lambda}$ hyperons has a larger magnitude than the one for $\Lambda$ hyperons, especially at $\sqrt{s_{NN}}=  7.7$GeV with SMSAH initial conditions. It opens a window to to probe the initial structure of QGP at the baryon-rich region through the local polarization of $\Lambda$ and $\bar{\Lambda}$

In  Fig.~\ref{fig:py_mode}, we plot the local polarization of $\Lambda$ hyperons along out-of-plane direction $P^y$ contributed from different sources in $20-50\%$ centrality at $\sqrt{s_{NN}} = 7.7, \;27,\;62.4 $GeV Au+Au collisions using AMPT and SMASH initial conditions.
We observe that the contribution from the thermal vorticity, $P^y_{\rm th}$, dominates over other sources for both AMPT and SMASH initial conditions.
The simulations with SMASH initial condition give a larger $P^y_{\rm th}$ than with the AMPT models, especially at $\sqrt{s_{NN}} =7.7$GeV.
For the polarization induced by SHE, $P^y_{\rm chem}$, the results from both AMPT and SMASH initial conditions are similar.
We note that in Fig.~\ref{fig:pz_mode} the polarization induced by SHE along the beam direction, $P^z_{\rm chem}$ has the similar behavior as the one induced by the shear tensor, $P^z_{\rm shear}$. In contrast, the slope of $P^y_{\rm chem}$ seems to be always opposite to the one of $P^y_{\rm shear}$, as shown in Fig.~\ref{fig:py_mode}.
Also similar to  Fig.~\ref{fig:pz_mode}, the polarization related to the fluid acceleration, $P^y_{\rm accT}$, is much smaller than other sources.

Next, let us focus on the collision energy dependence of the local polarization along the out-of-plane direction induced by different sources. Similar to the vorticity derived from other models \citep{Deng:2016gyh,Wei:2018zfb,Deng:2020ygd,Huang:2020xyr}, $P^y_{\rm th}$ increases when the collision energies decrease for both initial conditions. As for the transverse polarization induced by SIP and SHE, $P^y_{\rm shear}$ and $P^y_{\rm chem}$ with AMPT initial condition first increase and then decrease when the collision energy decreases. With SMASH initial condition, we find that $P^y_{\rm chem}$ increases monotonically with decreasing collision energy, while $P^y_{\rm shear}$ first increases and then flips its sign when the collision energy decreases.
Such behavior is different from the results in Ref. \citep{Fu:2022myl} due to the choice of initial conditions and parameters.
The most important observation in Fig. \ref{fig:py_mode} is that the SIP contribution, $P^y_{\rm shear}$, is sensitive to the initial conditions.
$P^y_{\rm shear}$ changes its sign at $7.7$GeV with SMASH initial condition, which is not observed for AMPT initial condition. Note that $P^y_{\rm shear}$ at $7.7$GeV with SMASH initial condition is also quite different
from the results obtained by other hydrodynamics simulations \cite{Fu:2022myl}.

In Fig.~\ref{fig:py_mode_total}, we present the total local spin polarization along out-of-plane direction $P^y$ of $\Lambda$ and $\bar{\Lambda}$ hyperons as a function of $\phi_p$.
The magnitude of
$P^y$ from both AMPT and SMASH initial conditions increases with decreasing collision energies. As shown in Fig.~\ref{fig:py_mode}, $P^y_{\textrm{th}}$ from SMASH initial condition is larger than the one from AMPT initial condition. The simulations with SMASH initial condition give a larger $P^{y}$ compare to AMPT initial condition. The difference between the magnitudes of $P^y$ from two initial condition models increases when the collision energies decrease.
At $27$ and $62.4$GeV, we also find that the polarization is smaller at in-plane direction than at out-of-plane direction.
On the other hand, at $7.7$GeV, the total local polarization $P^y$ from AMPT initial conditions is almost independent on the azimuthal angle $\phi_p$.

\subsection{Baryon diffusion dependence} \label{subsec:baryon}

As is known, the effect of baryon diffusion is crucial at finite net baryon density region. In this subsection, we study the effect of baryon diffusion on $\Lambda$ polarization.

In Figs.~\ref{fig:pz_mode_cb} and
\ref{fig:pz_total_cb},
we change the baryon diffusion coefficient to be $C_B=1.2$ in the longitudinal local polarization $P^z$ and compare to the results shown in
Figs. \ref{fig:pz_mode} and \ref{fig:pz_polar_total}.
In Figs.~\ref{fig:pz_mode_cb},
we find that the SIP, $P^z_{\textrm{shear}}$, and the polarization induced by the fluid acceleration, $P^z_{\textrm{accT}}$, in the case of $C_B=1.2$ are almost the same as those with $C_B =0$.
On the other hand, the polarization along the beam direction induced by the thermal vorticity, $P^z_{\textrm{th}}$, and SHE, $P^z_\textrm{chem}$, are enhanced when $C_B$ increases. Such enhancement is prominent at low energy collisions and in the simulations with SMASH initial condition.

We present the total local polarization along the beam direction for $\Lambda$ and $\overline{\Lambda}$ hyperons in
Fig.~\ref{fig:pz_total_cb} with $C_B=0$ and $1.2$.
We find that the magnitude of total $P^z$ for both $\Lambda$ and $\overline{\Lambda}$ hyperons from AMPT initial condition slightly decrease when $C_B$ increases at $\sqrt{s_{NN}} = 27,\;62.4 $GeV. On the other hand, the total $P^z$ from SMASH initial condition is very sensitive to the value of $C_B$.
Interestingly, we observe that the total local polarization $P^z$ for $\Lambda$ hyperons from SMASH initial condition will change its ``sign" at $7.7$GeV when $C_B=1.2$. This is due to the enhancement of the polarization induced by SHE, $P^z_\textrm{chem}$, from the baryon diffusion effect as shown in Fig. \ref{fig:pz_mode_cb}.

At last, we also plot the local polarization along $y$ direction for $\Lambda$ hyperons, $P^y$, with different $C_B$ in Fig.~\ref{fig:py_mode_cb}. The baryon diffusion effect on $P^y$ induced by different sources with both AMPT and SMASH initial conditions are almost negligible. This means that the local transverse polarization of $\Lambda$ hyperons is only very sensitive to initial conditions as shown above. Thus it should provide a very good probe to the initial states of relativistic heavy-ion collisions.

\subsection{Global polarization} \label{subsec:global}

\begin{figure}[tb]
\includegraphics[scale=0.42]{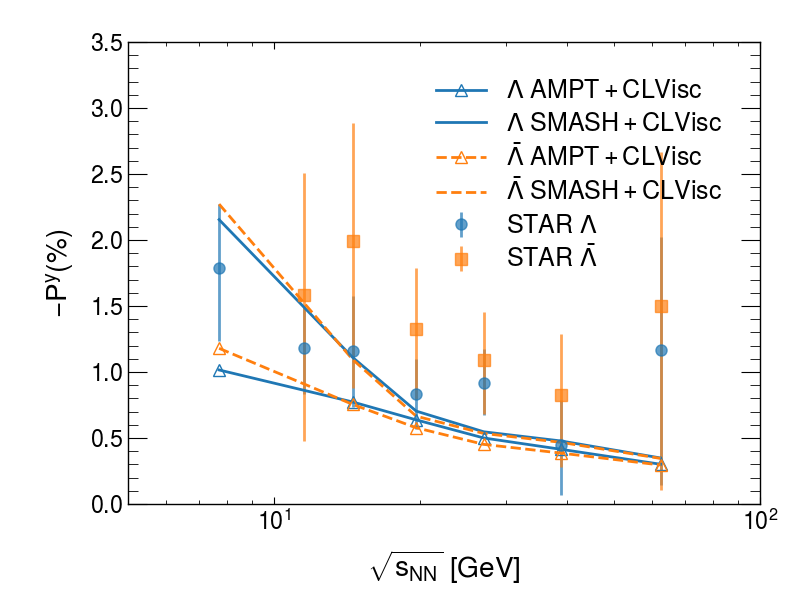}
\caption{The  global polarization $P^y$ as a function of collision energies $\sqrt{s_{NN}}$ for $\Lambda$ and $\overline{\Lambda}$ hyperons in $20-50\%$ centrality at Au+Au collisions. The experimental data are from the STAR measurements \citep{STAR:2021beb} and scaled by 0.877 .  We choose $p_T \in [0.5,3.0]$, $y\in [-1,1] $ and $C_B=0$.
}
\label{fig:py_global_ampt_SMASH}
\end{figure}

In Fig.~\ref{fig:py_global_ampt_SMASH}, we show the global polarization of $\Lambda$ and $\bar{\Lambda}$  hyperons along the out-of-plane direction as a function of collision energies in the mid-rapidity region at $20-50\%$ centrality Au+Au collisions.

Our numerical results agree with the STAR measurements very well.
The simulations with both both AMPT and SMASH initial conditions
show that the total polarization along out-of-plane direction $P^y$ increases when the collision energies decreases.
We numerically check that the global polarization is mainly induced
by thermal vorticity after integrating over the azimuthal angle $\phi_p$.
At the low energy collision, e.g. when $\sqrt{s_{NN}} \leq 27$GeV,
the polarization computed from SMASH initial condition is much larger
than the one from AMPT initial condition. This difference comes from the effect of
finite nuclear thickness, which is included in the SMASH model.
Interestingly, we also find that the polarization of $\overline{\Lambda}$ hyperons is not always larger than the one of $\Lambda$ hyperons, which is different with other studies \citep{Ryu:2021lnx}. Our results implies that there are competitions
between the finite baryon chemical potential effect and the production times of $\Lambda$ and $\bar{\Lambda}$ hyperons.

\section{Conclude and discussion}
\label{summary}

We have systematically studied the local and global polarization of
$\Lambda$ and $\bar{\Lambda}$ hyperons at RHIC-BES energies in the framework of
the (3+1)-dimensional CLVisc hydrodynamic model with AMPT and SMASH initial conditions.
In this work, we concentrate on the three aspects, the local polarization induced by SHE, the role of initial condition and the dependence of baryon diffusion to the local polarization.

We consider two different initial conditions for our simulations based on AMPT and SMASH models. We emphasize that the finite thickness effects of nucleus have already been encoded in SMASH. This effect plays an important role at baryon rich region of collisions. The polarization pseudo vector is given by the modified Cooper-Frye formula, in which the axial current in phase space is derived from QKT. Then, the polarization can be decomposed as the polarization induced by thermal vorticity, shear tensor, the gradient of baryon chemical potential over temperature, the fluid acceleration and electromagnetic fields. For simplicity, we neglect the contribution from electromagnetic fields. In this work, we compute the local and global polarizations in $20-50\%$ centrality at $\sqrt{s_{NN}}=7.7,\; 27,\; 62.4$GeV Au+Au collisions.

In Sec. \ref{subsec:local}, we concentrate on the local polarization induced by SHE and the role of initial conditions.
At $\sqrt{s_{NN}}=27,\; 62.4$GeV, the simulations with both AMPT and SMASH initial conditions show negligible contribution from the polarization induced by SHE.
At $\sqrt{s_{NN}}=7.7$GeV, the polarization induced by SHE, $P^z_\textrm{chem}$, from the simulations with AMPT initial condition, gives a sizeable contribution and can even flip the sign of the total local polarization along the beam direction $P^z$, while $P_{\text{chem}}^{z}$ from SMASH initial condition at that collision energy region
is negligible.
The polarizations along the out-of-plane direction induced by SHE, $P_{\text{chem}}^{y}$, from AMPT
initial condition first increases and then decrease when the collision
energy decreases, while the $P_{\text{chem}}^{y}$ form SMASH initial
condition increases monotonically with decreasing collision energy.

We also discuss the local polarizations induced by thermal vorticity and shear tensor.
$P_{\text{th}}^{z}$ and $P_{\text{shear}}^{z}$ from the
AMPT initial condition are similar to those from SMASH initial condition because of the similar radial flows generated from two initial conditions.
Along the out-of-plane direction, we observe that $P_{\text{th}}^{y}$ dominates the total local polarization.
$P_{\text{th}}^{y}$ from SMASH initial condition is larger
than that from AMPT initial condition, especially at $\sqrt{s_{NN}}=7.7\text{GeV}$.
In particular, $P_{\text{shear}}^{y}$ from SMASH initial condition change its trend at $\sqrt{s_{NN}}=7.7\text{GeV}$ compared to the results at $\sqrt{s_{NN}}=27,\; 62.4$GeV. Such behavior has not been observed for AMPT initial condition.
We also notice that the slope of $P_{\text{chem}}^{z}$ is similar to $P_{\text{shear}}^{z}$, while the slope of $P_{\text{chem}}^{y}$ is opposite to the $P_{\text{shear}}^{y}$.

Next, we have studied the dependence on baryon diffusion coefficients $C_B$ in Sec. \ref{subsec:baryon}. We find that $P^z_\textrm{th}$ and  $P^z_\textrm{chem}$ are sensitive to $C_B$ especially at $7.7$GeV with SMASH initial condition. The results from SMASH initial condition at $7.7$GeV collision show that the total $P^z$ flip its sign when $C_B=1.2$. On the other hand, the effects of finite $C_B$ to the polarization along the out-of-plane direction, $P^y$, are negligible.

We also plot the global polarization of $\Lambda$ and $\overline{\Lambda}$ hyperons  as a function of collision energy in Sec. \ref{subsec:global}. Our results agree with the STAR measurement. Interestingly, we find that the global polarization of $\overline{\Lambda}$ hyperons is not always larger than the one of  $\Lambda$ hyperons due to the competing effects from the finite baryon chemical potential and the production times of $\Lambda$ and $\bar{\Lambda}$ hyperons.

We conclude that the initial conditions and baryon diffusion are crucial to describe both the local and global polarization. The SHE contributions also play a role to both local and global polarizations.
The strong dependences of the initial conditions and the baryon diffusion imply the uncertainties in the theoretical frameworks.
Since polarization depends on more detailed structures of the QGP, such as thermal vorticity, shear tensor and the gradients of baryon chemical potential, compared to other collective phenomena, future studies should be able to provide better understanding of the novel properties of the QGP in relativistic heavy-ion collisions.

\begin{acknowledgments}
We thank Bao-chi Fu, Long-Gang Pang, Yi Yin , Hui-chao Song, and Xin-Li Sheng for helpful discussions.
This work is supported in part by Natural Science Foundation of China (NSFC) under Grants No. 11775095, No. 11890710, No. 11890711, No. 11935007, No. 12075235 and No. 12135011.
Some of the calculations were performed in the Nuclear Science Computing Center at Central China Normal University (NSC$^3$), Wuhan, Hubei, China.
\end{acknowledgments}

\bibliographystyle{h-physrev5} %reference style
\bibliography{refs,qkt-ref}   % data for reference

\end{document}